\documentclass[aps,floatfix,twocolumn,showpacs,pra]{revtex4}
\usepackage{graphicx,bm}
\usepackage{amsmath}
\usepackage{amssymb}

\begin{document}
\title{Coherence time of a Bose-Einstein condensate}
\author{A. Sinatra}
\affiliation{Laboratoire Kastler Brossel,
Ecole Normale Sup\'erieure, UPMC and CNRS,
24 rue Lhomond, 75231 Paris Cedex 05, France}
\author{E. Witkowska}
\affiliation{Institute of Physics, Polish Academy of Sciences,
Aleja Lotnik\'ow 32/46, 02-668 Warszawa, Poland }
\author{Y. Castin}
\affiliation{Laboratoire Kastler Brossel, 
Ecole Normale Sup\'erieure, UPMC and CNRS,
24 rue Lhomond, 75231 Paris Cedex 05, France}

\begin{abstract}
Temporal coherence is a fundamental property of macroscopic quantum systems, such as lasers in optics and Bose-Einstein condensates in atomic gases
and it is a crucial issue for interferometry applications with light or matter waves. Whereas the laser is an ``open" quantum system,
ultracold atomic gases are weakly coupled to the environment and may be considered as isolated. 
The coherence time of a condensate is then {\it intrinsic} to the system and its derivation is out of the frame of laser theory.
Using quantum kinetic theory, we predict that the interaction with non-condensed modes gradually smears out the condensate phase, with a variance growing as 
${\cal A} t^2+{\cal B} t+{\cal C}$ at long times $t$, and we give a quantitative prediction for ${\cal A}$, ${\cal B}$ and ${\cal C}$. 
Whereas the coefficient ${\cal A}$ vanishes for vanishing energy fluctuations in the initial state, the coefficients ${\cal B}$ and ${\cal C}$ are remarkably insensitive 
to these fluctuations. The coefficient ${\cal B}$
describes a diffusive motion of the condensate phase that sets the ultimate limit to the condensate coherence time. We briefly discuss the possibility to observe
the predicted phase spreading, also including the effect of particle losses.
\end{abstract}


\maketitle

\section{Introduction}
\label{sec:intro}
Bose-Einstein condensation eventually occurs in a bosonic system, if one reduces the temperature at a fixed density. It is characterized by the macroscopic occupation of the lowest single particle energy mode and by the onset of long range coherence both in time and space.
Initially predicted by Einstein for an ideal Bose gas in 1924, it has now been observed in a wide range of 
physical systems: in liquid helium \cite{Kapitza,Allen}, in ultracold atomic gases \cite{Cornell,Ketterle}, and in a variety of condensed matter systems such as 
magnons in anti-ferromagnets \cite{magnonsBEC}, and exciton polaritons in microcavities \cite{polaritonsBEC}. 
Among all these systems, ultracold atomic gases offer an unprecedented control on experimental parameters and allow very precise
measurements as is custom in atomic physics.
Experimental investigation of time coherence in condensates began right after their achievement in the laboratory \cite{JILA,Kasevich,Bloch} and 
the use of condensates in atomic clocks or interferometers is currently a cutting-edge subject of investigation
 \cite{BEC_precision,Shin,Ketterle1D,Oberthaler}. Therefore a crucial issue is to determine the ultimate limits on the coherence time of these systems.
Unlike lasers and most solid state systems in which condensation has been observed, ultracold atomic gases are weakly coupled to their environment. The intrinsic coherence time of a condensate is then due to its interaction with the non-condensed modes in an ideally {\it isolated} system,
which makes the problem unique and challenging. 
For the one dimensional quasi-condensate a theoretical treatment exists \cite{Demler} that was successfully 
compared with experiment \cite{Widera,Schmidtmayer}. In a true three dimensional condensate, the problem was solved in \cite{Beliaev} at zero temperature
while until now it has been still open at non-zero temperature.

As it is known since the work of Bogoliubov \cite{Bogoliubov}, the appropriate starting point for the description of a weakly interacting degenerate Bose gas is that of a weakly interacting gas of quasi-particles: the Bogoliubov excitations. The interactions among these quasiparticles shall play the main role in our problem. They
have to be included in the formalism in a way that fulfills the constraint of energy conservation, a crucial point for an isolated system. 
A first set of works addressed the problem of phase coherence in condensates using open-system approaches in analogy with the laser
\cite{Zoller,Graham,Graham2}: diffusive spreading of the condensate phase was predicted. These works however are not to be considered as quantitative,
due to the fact that a simplified model is used in \cite{Zoller},
and due to an approximate expression
of the condensate phase derivative in \cite{Graham,Graham2}.
Moreover, lacking the constraint of energy conservation, 
these approaches neglect some long time correlations among 
Bogoliubov excitations
that are responsible for a ballistic spreading in time of the condensate phase as shown in \cite{Kuklov,PRA_Super} using many-body approaches.
Unfortunately the final prediction in \cite{Kuklov} does not 
include the interactions among Bogoliubov modes:
the Bogoliubov excitations then do not decorrelate in time, 
the prediction quantitatively disagrees 
with quantum ergodic theory \cite{PRA_Super},
and no diffusive regime for the condensate phase is found.
Finally, the ergodic approach in \cite{PRA_Super}, while giving the correct
ballistic spreading of the phase, cannot predict a diffusive term.

As we now explain, quantum kinetic theory allows 
to include both energy conservation and quasi-particle interactions,
and gives access to both the ballistic and the diffusive behavior
of the phase.
To be as general and as simple as possible,
we consider a homogeneous gas in a box of volume $V$ with periodic boundary conditions. The condensate then forms in the plane wave with wave vector ${\bf k}={\bf 0}$. 
The total number of particles is fixed to $N$ and the density is $\rho=N/V$.
Let us consider the phase accumulated by the condensate during a time interval $t$: $\hat{\varphi}(t)=\hat{\theta}(t)-\hat{\theta}(0)$ where $\hat{\theta}$ is the condensate phase
operator
\cite{note_phase}.
Due to the interactions with the Bogoliubov quasi-particles, the accumulated condensate phase will not be exactly the same in each realization of the experiment.
We say that the phase fluctuates and spreads out in time or that the variance $\rm{Var} \, {\hat{\varphi}}(t)$ is an increasing function of time.
In presence of energy fluctuations in the initial state, the variance of the phase grows quadratically in time as already mentioned
\cite{Kuklov,PRA_Super}. Quantitatively this may be seen as follows: for $t\to\infty$, $\hat{\varphi}(t)\sim - \mu(E)t/\hbar$ where $\mu(E)$ is the chemical potential which depends only on the energy of the isolated system
\cite{PRA_Super}. By linearizing $\mu(E)$ around the average energy $\bar{E}$ 
for small relative energy fluctuations, one finds
\begin{equation}
{\rm{Var}} \, \hat{\varphi} (t) \sim \left( \frac{d\mu}{dE}\right)_{E=\bar{E}}^2 {\rm{Var}}\, E \; \frac{t^2}{\hbar^2} \,. \label{eqs:varphi}
\end{equation}
This ballistic spreading in time of the phase is comparable to that 
of a group of cars traveling with different speeds.
What happens if one reduces ideally to zero the energy fluctuations in the initial state ? We will show that the condensate phase will still spread but more slowly, with a diffusive motion. A precise calculation of the diffusion coefficient of the condensate phase in different experimental conditions, with or without fluctuations in the initial energy is the main goal of this paper.

The paper is organized as follows. The most important section is the overview section \ref{sec:main_result}: there we present the main results 
of the paper that we test against classical field simulations, and we indicate two possible schemes to observe them experimentally with cold atoms. 
Further precisions and all the technical details are given in the subsequent sections.
Starting from kinetic equations in section \ref{sec:equations_cinetiques}, that we 
linearize and solve in section \ref{sec:sole}, we obtain explicit results for the phase variance in section \ref{sec:results}. We discuss the effect of losses
in section \ref{sec:losses} and we conclude in section \ref{sec:conclusions}.

\section{Overview and Main Results}
\label{sec:main_result}
\begin{figure}[htb]
\centerline{\includegraphics[width=7cm,clip=]{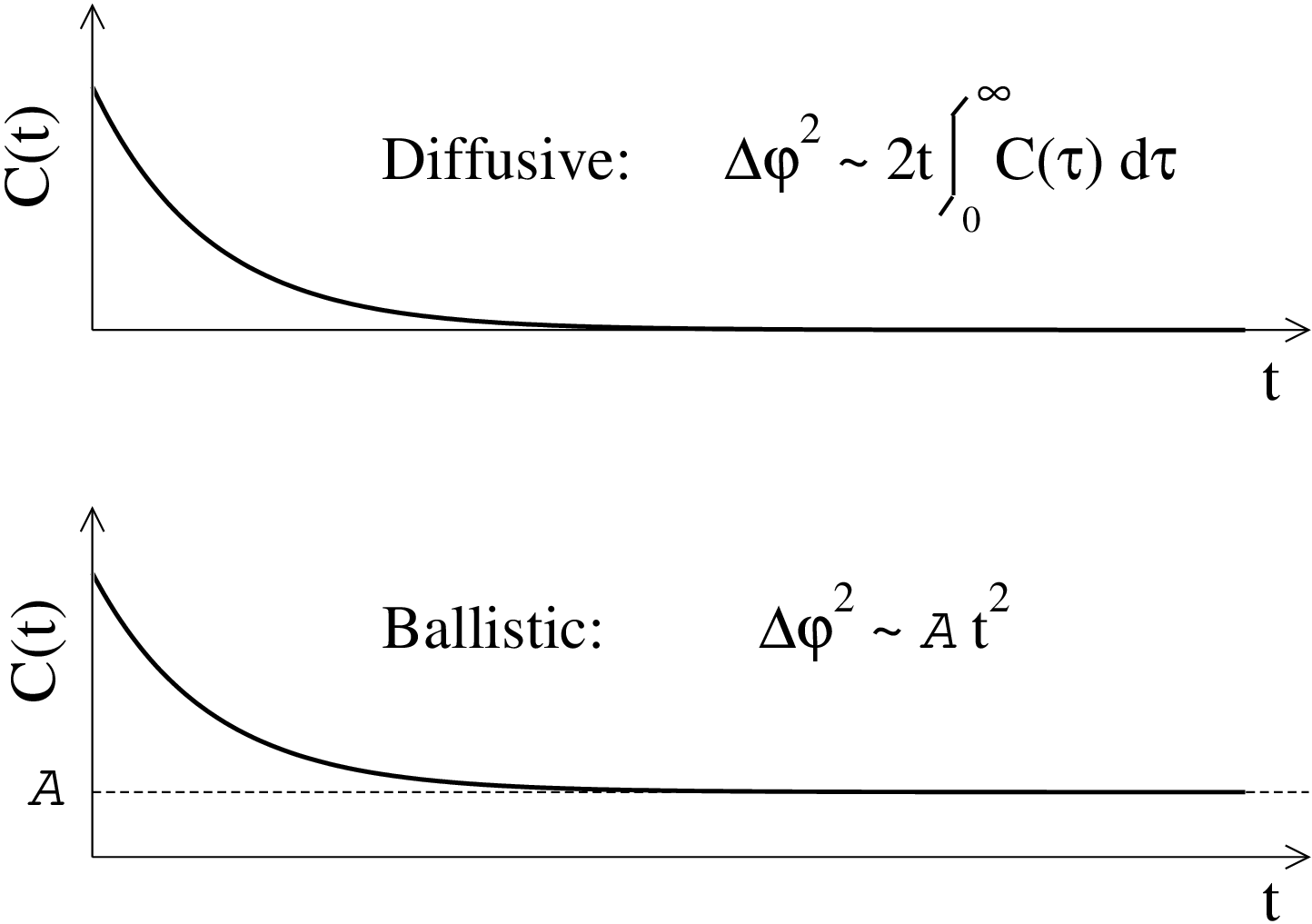}}
\caption{Schematic view of the correlation function of the condensate phase derivative $C(t)$. If $C(t)$ tends to zero fast enough for $t\to \infty$, the phase
spreading is diffusive. 
If $C(t)$ tends to a constant ${\cal A}$, the phase spreading is ballistic.
\label{fig:corr_thdot}}
\end{figure}

For a low temperature gas $T\ll T_c$ the temporally coarse-grained derivative of the condensate phase can be expressed in terms of the numbers $\hat{n}_{\bf k}$ of quasi-particles of wave vector ${\bf k}$ \cite{PRA_Super}
\begin{equation}
\dot{\hat{\varphi}} \simeq -\frac{\mu_0}{\hbar} -\sum_{{\bf k}\ne{\bf 0}} A_{\bf k} \hat{n}_{{\bf k}} 
\label{eq:varphidot}
\end{equation}
where the constant term $\mu_0$ is the ground state chemical potential of the gas and $A_{{\bf k}}=\frac{g}{\hbar V} (U_k+V_k)^2$. The coupling constant  $g$
for interactions between cold atoms is linked to the 
$s$-wave scattering length $a$ by $g=4\pi \hbar^2 a/m$, $m$ being the atom mass, and $U_k$, $V_k$ are the coefficients of the usual Bogoliubov modes:
\begin{equation}
U_k + V_k = \frac{1}{U_k-V_k} = \left(\frac{\hbar^2 k^2/(2m)}
{2\rho g +\hbar^2 k^2/(2m)}\right)^{1/4}\,.
\end{equation}
As a consequence of (\ref{eq:varphidot}), the variance of the condensate phase is determined by the correlation functions of the Bogoliubov quasiparticle
numbers $\hat{n}_{{\bf k}}$. 
Let $C(t)$ be the time correlation function of the condensate phase derivative \cite{note_corr_symm}:
\begin{equation}
C(t)  = \langle \dot{\hat{\varphi}}(t)  \dot{\hat{\varphi}}(0) \rangle -  \langle \dot{\hat{\varphi}}(t) \rangle  \langle  \dot{\hat{\varphi}}(0) \rangle\,.
\label{eq:C(t)}
\end{equation} 
By integrating formally $\dot{\hat{\varphi}}(t)$ over time and using time translational invariance:
\begin{equation}
{\rm Var}\,\hat{\varphi}(t)= 2t \int_0^t C(\tau)d\tau - 2\int_0^t \tau C(\tau) d\tau \,.
\label{eq:varphi}
\end{equation}
From equation (\ref{eq:varphi}) we see that two possible cases can occur. If $C(\tau)$ is a rapidly decreasing function of $\tau$ so that the integrals converge for $t\to \infty$, the variance of the phase will  grow linearly in time for long times and the condensate phase undergoes a diffusive motion with
a diffusion coefficient
\begin{equation}
D=\int_0^\infty C(\tau) d\tau \,.
\label{eq:DandC}
\end{equation}  
If $C(\tau)$ tends to a non zero constant value for $\tau \to \infty$, the phase variance grows quadratically in time and the phase undergoes a ballistic spreading.
 The two different scenarios are illustrated in Figure \ref{fig:corr_thdot}.
To describe the evolution of the quasiparticles number fluctuations $ \delta \hat{n}_{{\bf k}}(t)= \hat{n}_{{\bf k}}(t)- \langle \hat{n}_{{\bf k}}\rangle$ we write
quantum kinetic equations \cite{kinetic_equations} that we linearize.
Introducing the vector $\vec{A}$ of components $A_{{\bf k}}$,
the vector $\vec{x}(t)$ of components $x_{\bf k}(t)$:
\begin{equation}
x_{{\bf k}}(t) = \sum_{{\bf k'}\ne{\bf 0}} A_{{\bf k}'} \langle \delta \hat{n}_{{\bf k}}(t) \delta \hat{n}_{{\bf k}'}(0) \rangle \,,
\label{eq:x}
\end{equation}
and the matrix $M$ of linearized kinetic equations, one has:
\begin{equation}
\dot{\vec{x}}(t)=M \vec{x}(t)\,.
\label{eqs:xdot}
\end{equation}
Knowing $\vec{x}(t)$ we can calculate the phase derivative correlation function as
\begin{equation}
C(t)=\vec{A} \cdot \vec{x}(t) \,.
\label{eq:C_x(t)}
\end{equation}
On the basis of these equations we get our main result, that is the asymptotic expression of the variance of the condensate accumulated phase at long times:
\begin{equation}
{\rm{Var}}\, {\hat{\varphi}}(t) \simeq {\cal A} t^2 + {\cal B} t + {\cal C}  \:\:\:\:\:\: {\rm for} \:\:\:\:\:\: t\to \infty \,.
\label{eqs:ABC}
\end{equation}
In what follows we give an explicit expression for the coefficients  ${\cal A}$, ${\cal B}$ and ${\cal C}$. 

\begin{figure}[hob]
\centerline{\includegraphics[width=8cm,clip=]{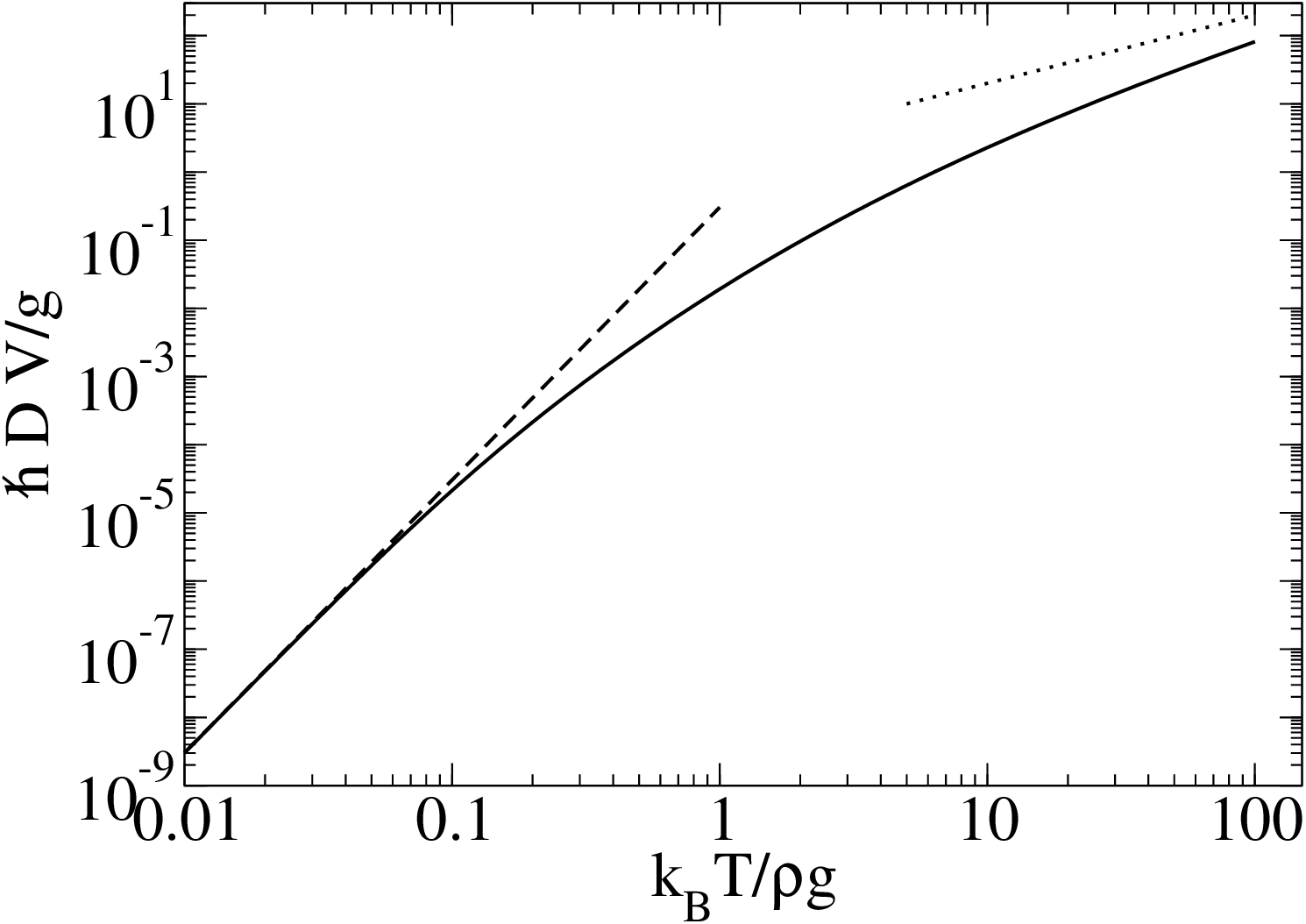}}
\caption{Rescaled diffusion coefficient of the condensate phase as a function of the rescaled temperature. Full line: 
numerical result from the solution of (\ref{eqs:D_final}). Dashed line: analytical prediction at low temperature: $y=0.3036 x^4$ (see Appendix
\ref{app:lowT}). Dotted line: 
approximate prediction of linear scaling at high temperature: $y \propto x$ (see Appendix \ref{app:highT}).
\label{fig:D_T}}
\end{figure}

\begin{figure}[t]
\centerline{\includegraphics[width=8.5cm,clip=]{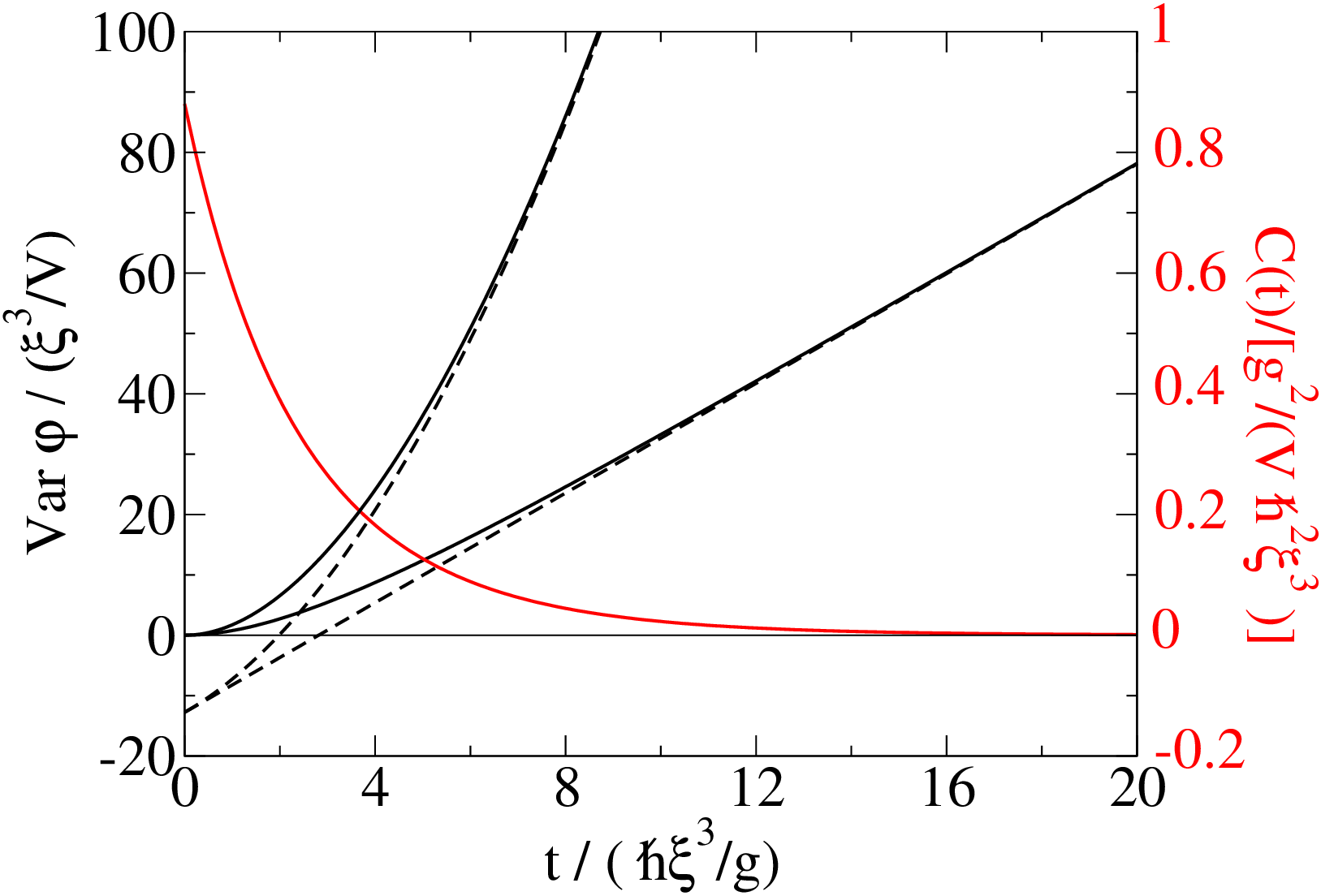}}
\caption{(Color online) Variance of the condensate accumulated phase as a function of time for $k_BT/\rho g=10$. Black full lines: ${\rm Var}\, \hat{\varphi} (t)$. 
Dashed lines: asymptotic behavior (\ref{eqs:ABC}). Red line (axis labels on the right): correlation function of the phase derivative $C(t)$. The upper curves for ${\rm Var}\,\hat{\varphi}(t)$ are obtained in presence of canonical ensemble energy fluctuations in the initial state. The lower curves, as well as
$C(t)$ correspond to the microcanonical ensemble where ${\cal A}=0$. In typical atomic condensates the healing length $\xi$ such that $\hbar^2/2m\xi^2=\rho g$ is at most in the $\mu$m range 
and the unit of time $\hbar \xi^3/g$ is at most in the ms range.
\label{figs:varphi}}
\end{figure}

The matrix $M$ has a zero frequency eigenvector $\vec{u}_0$. We then split the correlation vector $\vec{x}$ into two components:
$\vec{x}=\gamma \vec{u}_0 + \vec{X}(t)$.
The component of  $\vec{x}$ along $\vec{u}_0$ is constant in time. If it is non zero, $C(t)$
does not decay to zero for $t\to \infty$ and
the phase variance will grow quadratically.
In our general formalism we can show that $\gamma$ is 
linked to energy fluctuations in the initial state
and we recover the result (\ref{eqs:varphi}) for the coefficient ${\cal A}$.
The remaining component $\vec{X}$ has zero mean energy.
For the linear coefficient ruling diffusive phase spreading we find ${\cal B}=2D$ with:
\begin{equation}
D=-\vec{A} \cdot M^{-1} \vec{X}(0) \,,
\label{eqs:D_final}
\end{equation}
and for the constant term
\begin{equation}
{\cal C}=-2 \vec{A} \cdot M^{-2} \vec{X}(0) \,.
\label{eqs:calC_final}
\end{equation}
Remarkably $\vec{X}$, and thus $D$ and ${\cal C}$, do not depend on the energy fluctuations of the initial state, up to second order in the relative
energy fluctuations. 
We find that, in the thermodynamic limit, 
the rescaled diffusion coefficient $\hbar D V/g$ is a universal 
function of $k_BT/\rho g$ that 
we show in Figure \ref{fig:D_T}. 
This universal scaling was also found in \cite{PRA_Micro} in the frame of a classical field model.
At low temperature, we have shown analytically that ${\hbar D V}/{g}$ scales as the fourth power of ${k_BT}/{\rho g}$, while at high
temperature the rescaled diffusion coefficient grows approximately linearly with  ${k_BT}/{\rho g}$ (we expect logarithmic corrections to this law).
As made evident by the rescaling, $D$ is proportional to the inverse of the system volume and thus vanishes in the thermodynamic limit. The same property holds
for ${\cal C}$, and also for ${\cal A}$ in the case of canonical ensemble energy fluctuations.
In Fig.\ref{figs:varphi} for the temperature value $k_BT/\rho g=10$ 
we show the correlation function of the condensate phase derivative 
$C(t)$, that we calculate integrating (\ref{eqs:xdot}) in time.
On the same plot we show the variance of the condensate phase as a function of time that is obtained from (\ref{eq:varphi}). 
The asymptotic behavior of  ${\rm Var} \, \hat{\varphi} (t)$  from equation (\ref{eqs:ABC})
is reached after a transient time that is typically the decay time of the correlation function $C(t)$.

\begin{figure}[t]
\centerline{\includegraphics[width=7.7cm,clip=]{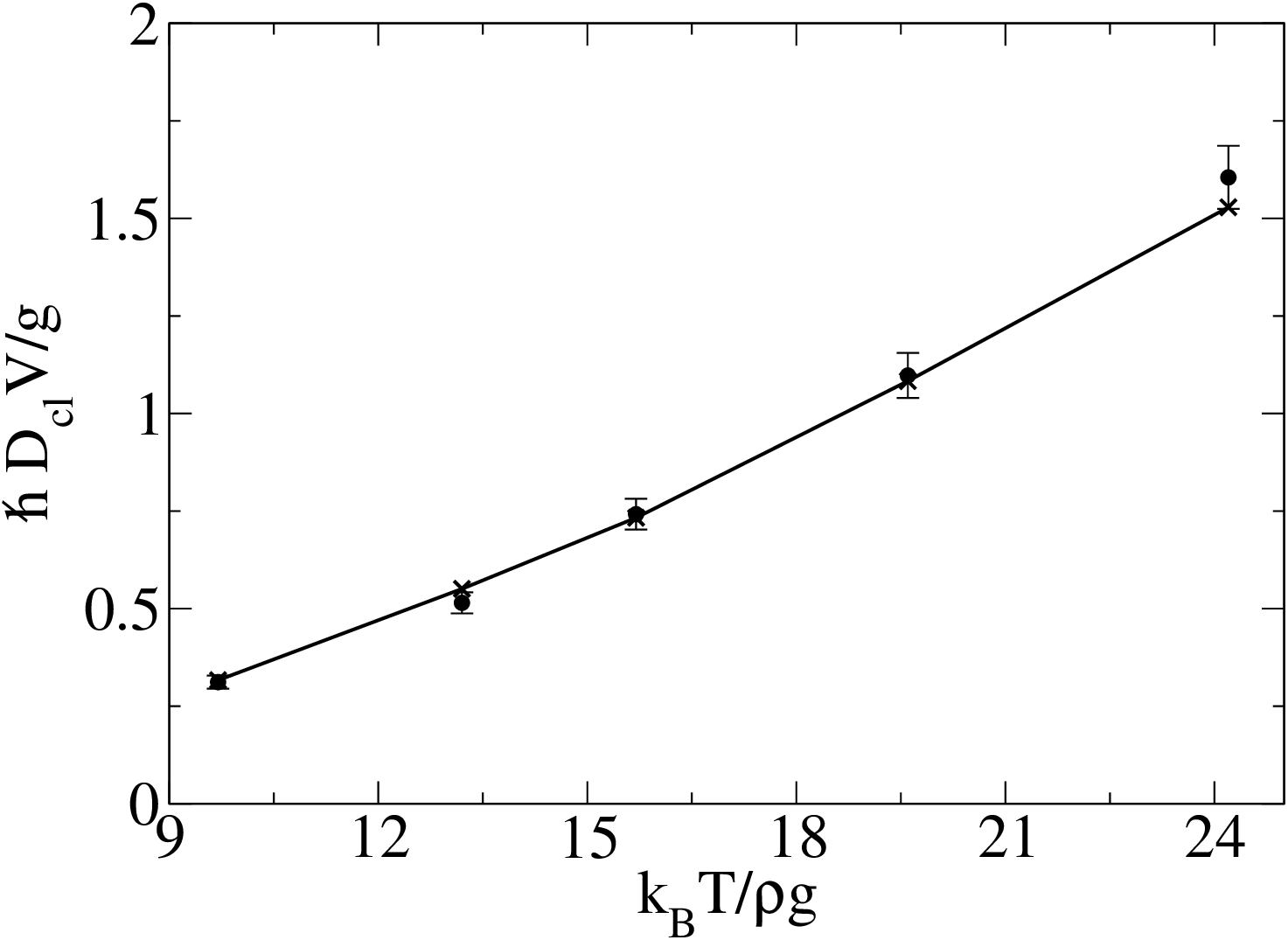}}
\caption{Diffusion coefficient $D_{cl}$ from the classical field theory on a lattice as a function of the temperature.
Crosses linked by a line: results from the classical version of our kinetic theory. Bullets with error bars: results from the classical field simulations with 1000 stochastic realizations in the microcanonical ensemble. 
In both curves there is a a cutoff at energy $k_BT$
\cite{note_cutoff}. 
\label{fig:comp_clf}}
\end{figure}

In the high temperature regime $k_BT/\rho g\gg1$, 
we were able to test our predictions against exact simulations
within a classical field model. 
In order to perform a quantitative comparison, we rephrased our kinetic theory for a classical field on a cubic lattice.
In both the classical kinetic theory and the classical field simulations we introduce an energy cut-off such that the maximum energy on the cubic lattice is 
of order $k_BT$ \cite{note_cutoff}.
We show the result of the comparison in Fig.\ref{fig:comp_clf}. 
As expected the numerical value of the diffusion coefficient $D_{cl}$ is different from the exact one given by the quantum theory and it depends in particular on the value of the cut-off. From the figure we find nevertheless a remarkable agreement between the classical kinetic theory and the classical field simulations. 

Our findings could have an immediate impact on present experiments with atomic condensates. Phase measurements have indeed already been successfully performed within two main schemes. 

The first scheme is out of equilibrium: starting from a condensate in a given internal state $a$, one applies two coherent short electromagnetic pulses separated by an evolution period for the condensate phase.
Each pulse transfers a fraction of the atoms into another internal state $b$. After the second pulse
one measures the number of atoms in state $b$.
In the original realization of this interferometric scheme \cite{JILA}, $\pi/2$ pulses were used which produce a strongly
out of equilibrium state of the system with a complex phase dynamics
\cite{Sinatra2000}. We propose to transfer only a tiny fraction of atoms in each of the two pulses, so that 
the depletion of the $a$ condensate and the interactions within $b$ atoms may be neglected. Moreover a spatial separation of $a$ and $b$ \cite{Philipp} or
a Feshbach resonance  \cite{Sengstock,Widera} may be used to suppress the
$a-b$ interactions. The ideal limiting case would be to transfer in $b$ a single atom which could be detected in a high finesse optical cavity
\cite{Reichel}. Using linear response theory one finds that the number of atoms in $b$ after the second pulse is proportional to
$N+ \Re e(e^{i\delta t_0}\langle \hat{a}_0^\dagger(t_0) \hat{a}_0(0) \rangle)$, where $\hat{a}_0$ is the condensate operator, $\delta$ is the detuning of the coherent pulses from the single atom $a-b$ transition and $t_0$ is the time interval between the two pulses. This signal is directly dependent on ${\rm Var}\, \hat{\varphi}(t)$. Indeed $|\langle \hat{a}_0^\dagger(t) \hat{a}_0(0) \rangle| \simeq N \exp[-{\rm Var} \, \hat{\varphi}(t)/2]$. 

The second scheme uses a symmetric atomic Josephson junction 
\cite{Oberthaler,Leggett},
in which one would cut the link between the two condensates by raising 
the potential barrier and measure the relative phase after an adjustable
delay time.
In this case an additional source of ballistic phase spreading is the partition noise proportional to the variance of the relative atom number. 
For homogeneous systems with canonical ensemble energy fluctuations 
on both sides of the Josephson junction, the ratio between this
undesired contribution and ${\cal A}t^2$ scales as $\xi_N^2 (\rho a^3)^{-1/2}/\tilde{\cal A}$ where $\xi_N^2$ is the number squeezing parameter of the Josephson junction, on the order of 0.35 in  \cite{Oberthaler}, and 
$\tilde{\cal A}={\cal A}/[ (\rho g/\hbar)^2(a^2 \xi/V)]$, where $\xi$ is the healing length, depends only on $k_BT/\rho g$ \cite{PRA_Super}. For 
$k_BT/\rho g=5$ one has $\tilde{\cal A}\simeq 150$ so that, for the typical value $(\rho a^3)^{1/2}=2.5\times10^{-3}$, the undesired contribution is smaller than ${\cal A}t^2$ \cite{t_tc_only}.

\section{Kinetic equations for the Bogoliubov excitations}
\label{sec:equations_cinetiques}
At low temperature $T\ll T_c$ we assume
that the state of the gas can be approximated 
by a statistical mixture of eigenstates of the Bogoliubov Hamiltonian $\hat{H}_{\rm Bog}$
\begin{equation}
\hat{H}_{\rm Bog} \equiv E_0+\sum_{{\bf k}\ne{\bf 0}} \epsilon_k \hat{n}_{{\bf k}}\,,
\label{eq:Hbog}
\end{equation}
where $E_0$ is the energy of the ground state.
The eigenstates of $\hat{H}_{\rm Bog}$
are Fock states $|\{ n_{\bf k} \}\rangle$  with well defined numbers of Bogoliubov quasiparticles. 
Whereas expectation values of stationary quantities are expected
to be well approximated by Bogoliubov theory,
this is no longer the case for two-time correlation functions.
This is physically quite clear for the correlation function of the Bogoliubov
mode occupation numbers $\hat{n}_{\bf k}$: whereas they never decorrelate
at the Bogoliubov level of the theory (they are conserved quantities
of $\hat{H}_{\rm Bog}$), they will experience some decorrelation for the full
Hamiltonian dynamics because of the interactions among Bogoliubov
quasi-particles, that are at the origin of the Beliaev-Landau processes.

For a given initial state of the system characterized by the occupation numbers $\{ n_{\bf k} \}=\{ n_{\bf k}(0) \}$ the time evolution,
beyond Bogoliubov approximation,
of the mean mode occupation numbers
\begin{equation}
n_{\bf q} (t) \equiv \langle \{ n_{\bf k}(0) \}|\hat{n}_{\bf q}(t)|\{ n_{\bf k}(0) \}\rangle 
\end{equation} 
can be described in terms of quantum kinetic equations of the form
\cite{kinetic_equations}: 
\begin{eqnarray}
\dot{n}_{{\bf q}} &=& -\frac{g^2 \rho}{\hbar \pi^2} \int d^3{k} \, \left\{
\left[ n_{{\bf q}} n_{{\bf k}}- n_{{\bf q}+{\bf k}} (1+ n_{{\bf k}} + 
n_{{\bf q}})\right]  \right. \nonumber \\
&& \left. \times \left({\cal A}_{k,q}^{|{\bf q}+{\bf k}|} \right)^2 \delta(\epsilon_q+\epsilon_k-\epsilon_{|{\bf q}+{\bf k}|}) \right\} \nonumber \\
&-&                 \frac{g^2 \rho}{2\hbar \pi^2} \int d^3{k}  \left\{
\left[ n_{{\bf q}}(1+n_{{\bf k}}+n_{{\bf q}-{\bf k}}) - n_{{\bf k}} n_{{\bf q}-{\bf k}} \right] \right. \nonumber \\
&& \left. \times \left({\cal A}_{k,|{\bf q}-{\bf k}|}^{q} \right)^2  \delta(\epsilon_k+\epsilon_{|{\bf q}-{\bf k}|}-\epsilon_q)  \right\}  \,.
\label{eq:kin}
\end{eqnarray}
In (\ref{eq:kin}) we have introduced in (\ref{eq:kin}) the coupling amplitudes
among the Bogoliubov modes:
\begin{eqnarray}
{\cal A}_{k,k'}^q &=& U_q U_{k} U_{k'} + V_q V_{k} V_{k'}  \nonumber \\
&+& (U_q+V_q)(V_{k} U_{k'} +U_{k} V_{k'}) \,.
\end{eqnarray}
Kinetic equations (\ref{eq:kin}) describe Landau and Beliaev processes in which
the mode of wave vector ${\bf q}$ scatters an excitation of wave vector ${\bf k}$ giving rise to an excitation
of wave vector ${\bf k}'$ (Landau damping),  the mode of wave vector ${\bf q}$ decays into an excitation of wave vector ${\bf k}$
and an excitation of wave vector ${\bf k}'$ (Beliaev damping), and inverse processes. In each process
the final modes have to satisfy energy and momentum conservation. Energy conservation is ensured by the delta distributions in (\ref{eq:kin}) where
$\epsilon_k$ is the Bogoliubov energy of the quasiparticle of wave vector ${\bf k}$, 
\begin{equation}
\epsilon_k = \left[\frac{\hbar^2 k^2}{2m}
\left(\frac{\hbar^2 k^2}{2m}+2\rho g\right)\right]^{1/2}.
\end{equation}

To calculate the correlation function $C(t)$, equations (\ref{eq:kin}) can be linearized for small deviations \cite{note_linearization},
and linear equations for the correlation functions $x_{{\bf q}}(t)$ can be obtained:
\begin{equation}
\dot{\vec{x}}=M \vec{x}\,.
\label{eq:xdot}
\end{equation}
To obtain $\dot{x}_{\bf q}$ from $\dot{n}_{\bf q}$, we connect expectation values in an initially considered Fock state to expectation values in the system state by an additional average. More details on the derivation
of (\ref{eq:xdot}), as well as the explicit form of the equations, which are in fact integral equations, are given in appendix \ref{sec:eqs_detail}. 
In particular, the matrix $M$ depends on the Bose occupation numbers
\begin{equation}
\bar{n}_q=\frac{1}{e^{\epsilon_q/k_B T}-1} \,.
\label{eq:nstat}
\end{equation}
The set of $\bar{n}_q$ constitutes a stationary solution of (\ref{eq:kin}),
with a temperature $T$ such that the mean energy of this solution
is equal to the mean energy of the system.
The classical version of kinetic equations that we used to test our results against classical field simulations (that are exact
within the classical field model)
are reported in appendix \ref{app:class}.

\section{Solution of the linearized equations}
\label{sec:sole}
The matrix $M$ is real and not symmetric. It has right and left eigenvectors $M \vec{u}_\lambda = m_\lambda \vec{u}_\lambda$,
${}^t\vec{v}_\lambda M = m_\lambda {}^t\vec{v}_\lambda$ satisfying $\vec{v}_\lambda \cdot \vec{u}_{\lambda'} = \delta_{\lambda\lambda'}$.
Due to the fact that the system is isolated during its evolution, $M$ has a pair of adjoint left and right eigenvectors with zero eigenvalue \cite{note_secteur}. 
Indeed for any fluctuation $\vec{\delta n}$,
introducing the vector $\vec{\epsilon}$ of components $\epsilon_k$, one has
\begin{equation}
\sum_{{\bf k}} \epsilon_k n_{{\bf k}} = {\rm{constant}} \to \sum_{{\bf k}} \epsilon_k \dot{\delta n}_{{\bf k}} =0 \to {}^t{\vec{\epsilon}}\, M {\vec{\delta n}} = 0 \,.
\label{eq:propr_u0}
\end{equation}
Let us denote $\vec{u}_0$ the right eigenvector of $M$ with eigenvalue 0 and $\vec{v}_0$ the corresponding left eigenvector.
One has from (\ref{eq:propr_u0}) $\vec{v}_0=\vec{\epsilon}$. On the other hand one can show that $\vec{u}_0=\vec{\alpha}$ \cite{note_u0} with : 
\begin{equation}
\alpha_\mathbf{k} = \frac{\epsilon_k \bar{n}_k (\bar{n}_k+1)}{\sum_{\mathbf q\neq \mathbf{0}} \epsilon_q^2 \bar{n}_q (\bar{n}_q+1)}.
\label{eq:def_alpha}
\end{equation}
It is useful to split the correlation vector $\vec{x}$ into a component parallel
to $\vec{\alpha}$ and a zero-energy component, that is a component
orthogonal to the vector $\vec{\epsilon}$:
\begin{equation}
\vec{x}=\gamma \vec{\alpha}+ \vec{X} \,.
\label{eq:splitx}
\end{equation}
For our normalization of $\vec{\alpha}$ one simply has
$\gamma = \vec{\epsilon}\cdot \vec{x}$.
From equations (\ref{eq:xdot}) and (\ref{eq:splitx}) we then obtain 
\begin{eqnarray}
\dot{\gamma}&=&0 \\
\dot{\vec{X}}&=&M \vec{X} \label{eq:Xdot} \,.
\end{eqnarray}
Under the assumption that $\vec{A} \cdot \vec{X}(\tau)=O(\tau^{-(2+\nu)})$ 
with $\nu > 0$ for $\tau \to \infty$, we obtain from (\ref{eq:varphi}) the asymptotic 
expression for the condensate phase variance:
\begin{equation}
{\rm Var} \, \hat{\varphi}(t) = {\cal A} t^2 + {\cal B} t + {\cal C} + o(1) \, \:\:\:\:\:\:{\rm for }  \:\:\:\:\:\: t\to \infty
\label{eq:genVarphi}
\end{equation}
with
\begin{eqnarray}
{\cal A}&=&\vec{A} \cdot \gamma \vec{\alpha}  \label{eq:calA} \\
{\cal B}&=& 2 \int_0^\infty d \tau \vec{A} \cdot \vec{X}(\tau) \label{eq:calB} \\
{\cal C}&=& -2 \int_0^\infty d \tau \, \tau \vec{A} \cdot \vec{X}(\tau) \,. \label{eq:calC}
\end{eqnarray}
As explained below, in the paragraph ``{\bf The correlation function $C(t)$}"
of section \ref{sec:results},
we have some reason to believe
that $\vec{A} \cdot \vec{X}(\tau)$ scales as $\tau^{-(2+\nu)}$ for large $\tau$ with $\nu=1$.

\section{Results for the phase variance}
\label{sec:results}
\noindent{\bf State of the system and quantum averages:}
In the general case, we assume that the state of the system is a statistical mixture of microcanonical states.
For any operator $\hat{O}$ one then has
\begin{equation}
\langle \hat{O} \rangle = \int dE \: P(E) \:  \langle \hat{O} \rangle_{\rm mc}(E) \,,
\label{eq:def_ave}
\end{equation}
where $\langle\ldots\rangle_{\rm mc}(E)$ is the microcanonical expectation
value for a system energy $E$.
Furthermore we make the hypothesis that the relative width of the energy distribution 
$P(E)$ is small. Formally, in the thermodynamic limit we assume
\begin{equation}
\frac{\sigma(E)}{\bar{E}} =O\left( \frac{1}{\sqrt{N}} \right) \hspace{0.5cm} {\rm for} \hspace{0.5cm} N \to \infty \,.
\label{eq:smallfluct}
\end{equation}
Besides microcanonical averages $\langle \hat{O} \rangle_{\rm mc}(E)$, we introduce canonical averages $\langle \hat{O} \rangle_{\rm can}(T)$ where the temperature $T$ is chosen such that $ \langle \hat{H}_{\rm Bog} \rangle_{\rm can}(T)=\langle \hat{H}_{\rm Bog} \rangle \equiv \bar{E}$. 
Useful relations among the quantum averages in the different ensembles are derived in Appendix \ref{app:quantumave}.

\noindent{\bf Quadratic term:}
First we calculate the quadratic term ${\cal A}$ of the condensate phase variance given in (\ref{eq:calA}). We introduce the ``chemical potential" operator  
 \begin{equation}
\hat{\mu} \equiv {\mu_0} + \hbar \sum_{{\bf k}\ne{\bf 0}} A_{\bf k} \hat{n}_{{\bf k}} 
\end{equation}
so that $-\hat{\mu}/\hbar=\dot{\hat{\varphi}}$ according to equation (\ref{eq:varphidot}).
The constant $\gamma$ appearing in  (\ref{eq:calA}) can then be expressed as
\begin{equation}
\gamma={\vec{\epsilon}}\cdot{\vec{x}}(0) =
\langle \left(\hat{H}_{\rm Bog}- \bar{E}\right) \: \hat{\mu} \rangle/\hbar
\end{equation}
so that, using (\ref{eq:def_ave}),
\begin{equation}
\gamma = \int dE \: P(E) \:  (E-\bar{E}) \langle \hat{\mu} \rangle_{\rm mc}(E)/\hbar \,.
\label{eq:compacte}
\end{equation}
We now expand the function $\langle \hat{\mu} \rangle_{\rm mc}(E)$ around its value for the average energy:
\begin{equation}
\langle \hat{\mu} \rangle_{\rm mc}(E)=\langle \hat{\mu} \rangle_{\rm mc}( \bar{E} ) + (E-\bar{E})\frac{d\langle \hat{\mu} \rangle_{\rm mc}}{dE}(\bar{E}) +\ldots
\label{eq:expansion_mu_mc}
\end{equation}
Inserting the expansion (\ref{eq:expansion_mu_mc}) in (\ref{eq:compacte})
one gets to leading order in the energy fluctuations:
\begin{equation}
\gamma \simeq \frac{d\langle \hat{\mu} \rangle_{\rm mc}}{dE}(\bar{E})
\frac{{\rm Var}\,E}{\hbar} \,.
\end{equation}
Using equation (\ref{eq:O_can_mc}) of Appendix \ref{app:quantumave} for $\hat{O}=\hat{\mu}$, we finally obtain
\begin{equation}
\gamma \simeq  \frac{\frac{d}{dT} \langle \hat{\mu} \rangle_{\rm can}}{\hbar\frac{d}{dT} \bar{E}} {\rm Var}\,E \,.
\end{equation}
According to (\ref{eq:calA}) we also need the value of $\vec{A} \cdot \vec{\alpha}$ that we can rewrite using 
(\ref{eq:C6}), (\ref{eq:C7}) as
\begin{equation}
\vec{A}\cdot \vec{\alpha} = \frac{\sum_{{\bf k}\ne{\bf 0}} A_{\bf k} \frac{d}{dT} \bar{n}_k }{\frac{d}{dT} \bar{E}} =
\frac{\frac{d}{dT} \langle \hat{\mu} \rangle_{\rm can}}{\hbar \frac{d}{dT} \bar{E}} \,.
\end{equation}
Finally
\begin{equation}
{\cal A}=\left( \frac{\frac{d}{dT} \langle \hat{\mu} \rangle_{\rm can}}
{\hbar\frac{d}{dT} \bar{E}}  \right)^2 {\rm Var}\,E \,.
\end{equation}
We then recover, by a different method and in a more general case, the main result of \cite{PRA_Super} for
super diffusive phase spreading when energy fluctuations are present in the initial state of the gas.

\noindent{\bf Linear term:}
The linear term ${\cal B}$ in (\ref{eq:calB}) represents a diffusion of the condensate phase with a diffusion coefficient
$D={\cal B}/2$. Integrating equation (\ref{eq:Xdot}) from zero to infinity and assuming $\vec{X}(\infty)=0$, we obtain
\begin{equation}
D=-\vec{A} \cdot M^{-1} \vec{X}(0)
\end{equation}
where the inverse of the matrix $M$ has to be understood in a complementary subspace to the kernel of matrix $M$,
that is in the subspace of vectors $\vec{x}$ satisfying $\vec{\epsilon} \cdot \vec{x}=0$. We can then write
\begin{equation}
D=-(P\vec{A}) \cdot M^{-1} \vec{X}(0)
\label{eq:D_fromM}
\end{equation}
where the matrix $P^\dagger$ projects onto this subspace in a parallel direction to $\vec{\alpha}$. This corresponds to a matrix $P$ given by
\begin{equation}
P_{\mathbf{k,k'}}=\delta_{\mathbf{k,k'}}-{\epsilon}_{\mathbf k} \alpha_{\mathbf{k'}} \, .
\end{equation}
As a consequence, one simply has
\begin{equation}
\vec{X}(0)=P^\dagger \vec{x}(0)=\vec{x}(0)-\vec{\alpha} \, \left(\vec{\epsilon}\cdot \vec{x}(0) \right) \,.
\end{equation}

\begin{figure}[htb]
\centerline{\includegraphics[width=8.5cm,clip=]{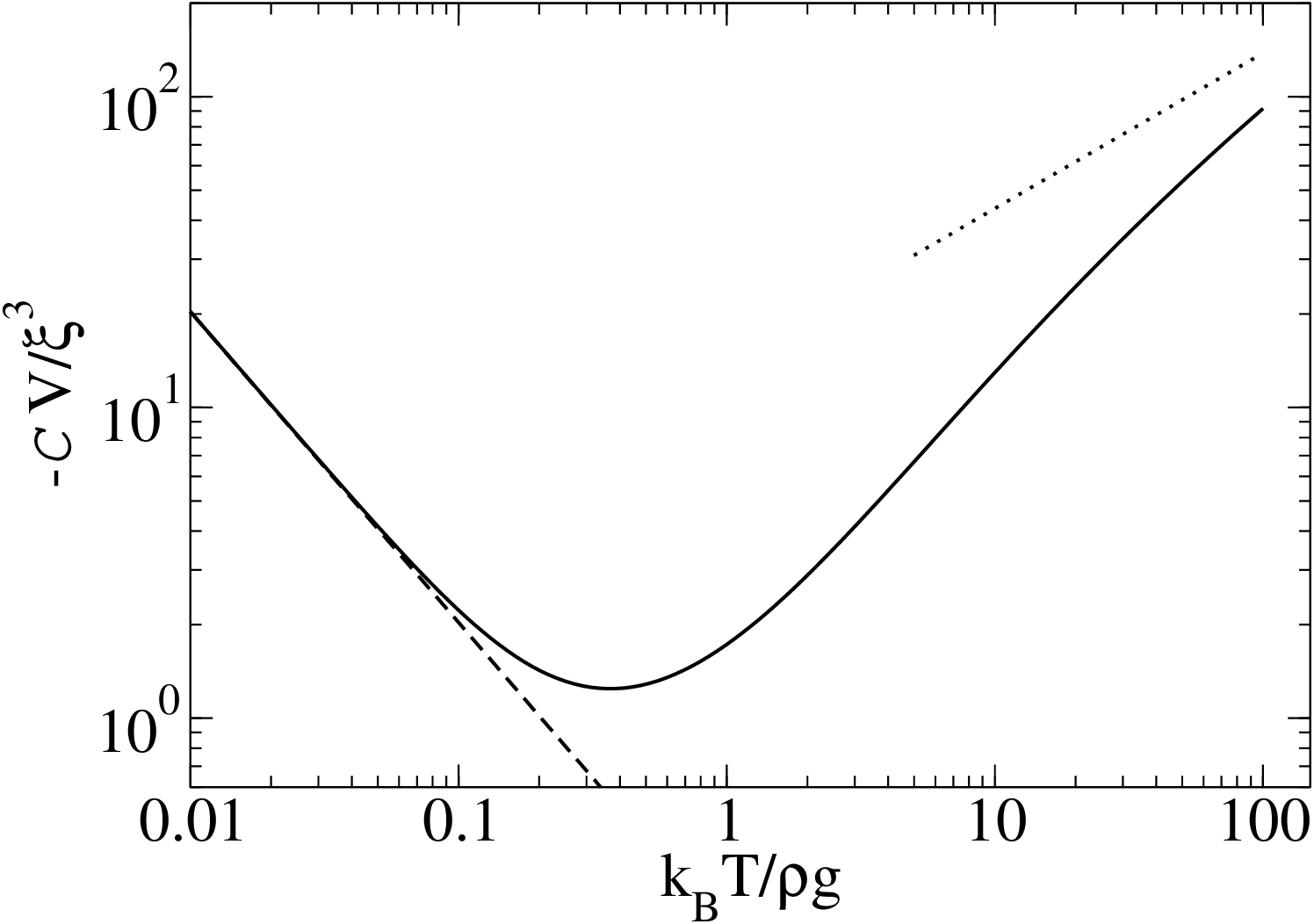}}
\caption{Constant ${\cal C}$ as a function of the rescaled temperature. Full line: 
numerical result from the solution of (\ref{eq:calC_final}). Dashed line: analytical prediction at low temperature: $y=0.2033 x^{-1}$ (see Appendix
\ref{app:lowT}). Dotted line naive prediction 
for the high temperature scaling: $y \propto x^{1/2}$ (see Appendix \ref{app:highT}).
\label{fig:c0_T}}
\end{figure}

We show here that $D$ does not depend on the width of the energy distribution $P(E)$ of the initial state. To this end
it is sufficient to show that the same property holds for $\vec{X}(0)$. We apply the relation (\ref{eq:aveO_final}) to $\hat{n}_{\bf k}$ 
and $\hat{n}_{\bf k} \hat{n}_{\bf k'}$ to obtain after some calculations
\begin{eqnarray}
\label{eq:dndn}
\langle \delta \hat{n}_{\bf k} \delta \hat{n}_{\bf k'} \rangle &=& 
\delta_{\bf k\, k'} \bar{n}_k(1+\bar{n}_k) 
\nonumber \\
&+&
(\eta-1)k_BT^2 
\frac{\left(\frac{d}{dT} \bar{n}_k\right) \left(\frac{d}{dT} \bar{n}_{k'}\right)}{\frac{d}{dT} \bar{E}} 
+\ldots
\end{eqnarray}
where the dots indicate terms giving higher order contributions
in the thermodynamic limit that will be neglected.
Here $\eta$ is the ratio of the variance of the system energy
to the energy variance in the canonical ensemble,
$\eta=\mbox{Var}\,E/\mbox{Var}_{\rm can}E$.
Eq.(\ref{eq:dndn}) shows that $\vec{x}(0)$ and hence $\vec{X}(0)$ 
are affine functions of $\eta$. $\vec{X}(0)$ can then be determined from its 
values in $\eta=0$ (microcanonical ensemble)
and $\eta=1$ (canonical ensemble):
\begin{equation}
\vec{X}(0)=\eta \vec{X}_{\rm can}(0) + (1-\eta)\vec{X}_{\rm mc}(0)\,.
\end{equation}
On the other hand one can show
explicitly for a large system
that $\vec{X}_{\rm can}(0)=\vec{X}_{\rm mc}(0)$ \cite{one_can_show}. As a consequence
\begin{equation}
\vec{X}(0) =\vec{X}_{\rm mc}(0)
\end{equation}
does not depend on $\eta$. 
Note that this relation extends to all positive times, $\vec{X}(t) = 
\vec{X}_{\rm mc}(t)$, since the matrix $M$ does not depend on the
energy fluctuations.

The expression of $\vec{X}_{\rm mc}(0)$ has been derived in \cite{PRA_Micro}.
Introducing the covariance matrix of Bogoliubov occupation numbers
\begin{equation}
Q_{\mathbf{k,k'}}^{\rm mc}(t)=\langle \delta \hat{n}_{\mathbf k}(t) \delta \hat{n}_{\mathbf k'}(0)   \rangle \,,
\end{equation}
one has in the microcanonical ensemble
\begin{equation}
\vec{X}_{\rm mc}(0)=Q^{\rm mc}(t=0)\vec{A}  \,.
\end{equation}

\begin{figure}[htb]
\centerline{\includegraphics[width=8.5cm,clip=]{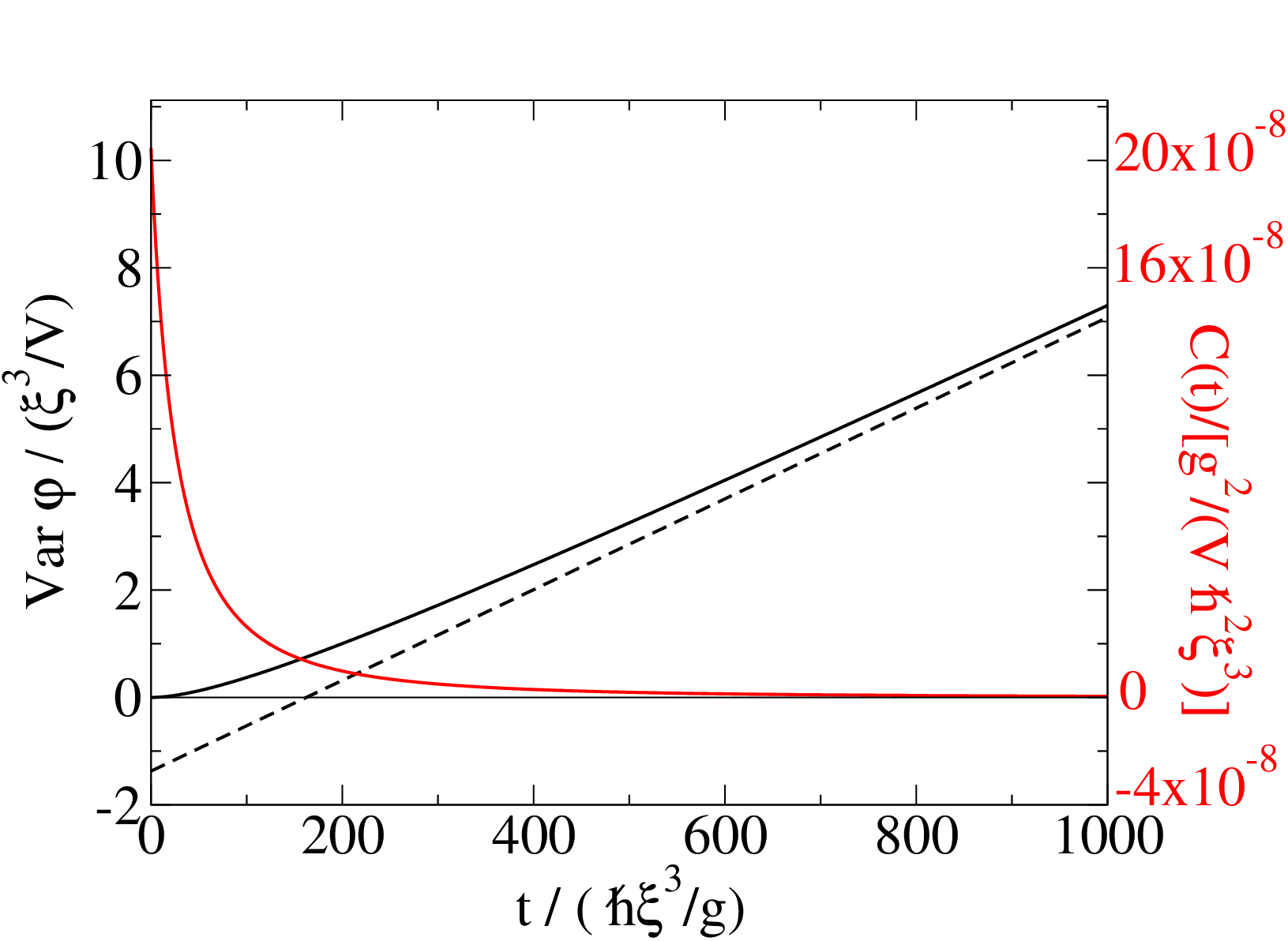}}
\centerline{\includegraphics[width=8.5cm,clip=]{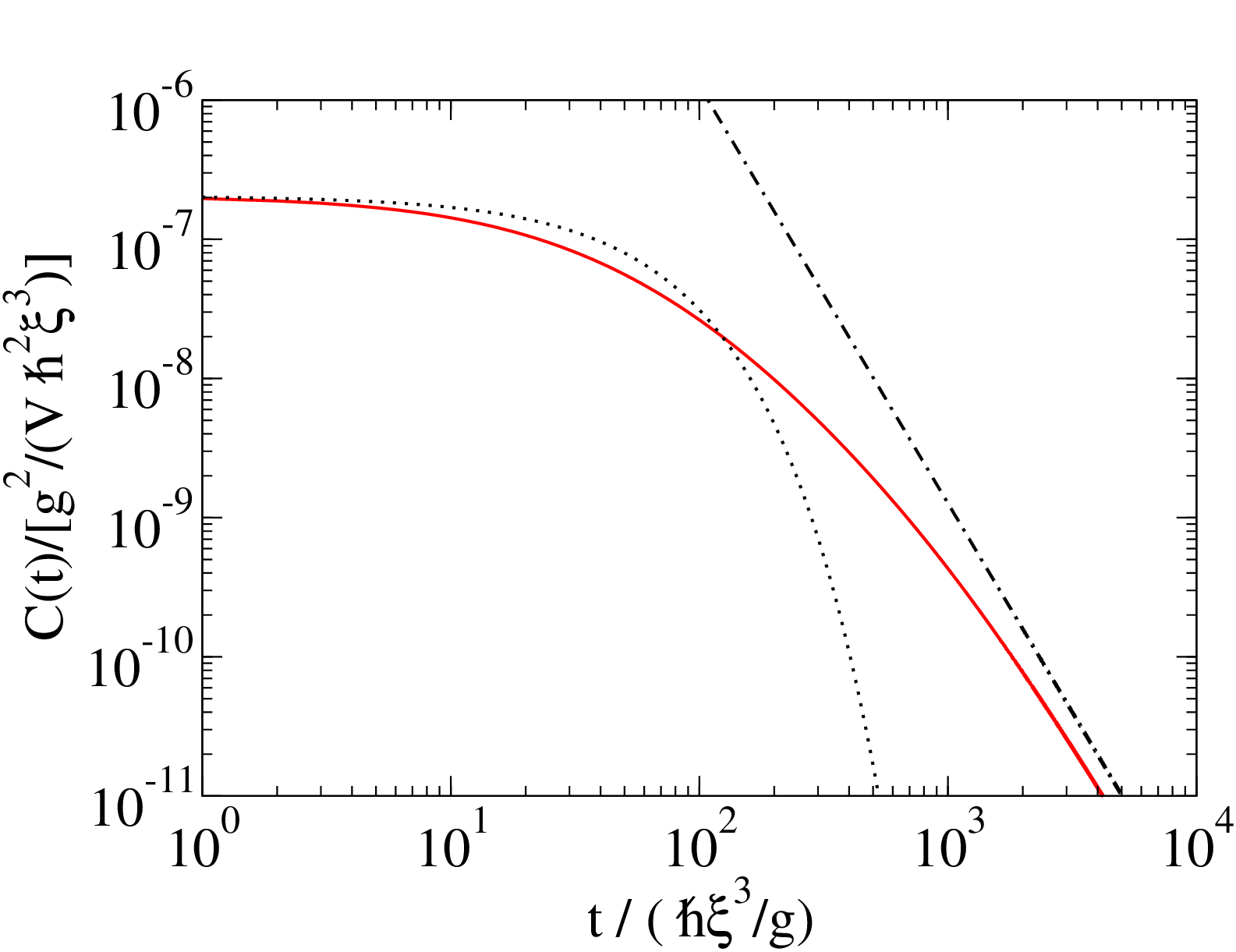}}
\caption{(Color online) {\bf Top:} For a system prepared in the microcanonical ensemble,
variance of the condensate accumulated phase as a function of time for $k_BT/\rho g=0.2$. Black full line: ${\rm Var}\, \hat{\varphi}$ obtained from (\ref{eq:varphi}).  Dashed line: asymptotic behavior (\ref{eq:genVarphi}). Red line: correlation function $C(t)$ defined in (\ref{eq:C(t)}).
{\bf Bottom}: Full red line: The same correlation function $C(t)$ in log-log scale.
Dotted line: Exponential function $f(t)=C(0)\exp(-t/\tau_c)$, 
where $\tau_c$ is defined in (\ref{eq:tauc}).
Dashed dotted line: law $y\propto x^{-3}$ predicted by the Gaussian model of \cite{PRA_Micro}.
$\xi$ is the healing length: $\hbar^2/2m\xi^2=\rho g$.
\label{fig:varphi}}
\end{figure}

As we showed in \cite{PRA_Micro}, for a large system,
the $t=0$ covariance matrix in the microcanonical ensemble can be obtained
by the one in the canonical ensemble by projection: 
\begin{equation}
Q^{\rm mc}(t=0) \simeq P^\dagger Q^{\rm can}(t=0) P \,,
\label{eq:initQ}
\end{equation}
where $Q^{\rm can}$ is the covariance matrix
in the canonical ensemble, that can be calculated using Wick's theorem
\begin{equation}
\label{eq:Qcan}
Q^{\rm can}_{\mathbf{k,k'}}(t=0)= \bar{n}_k (\bar{n}_k+1) \delta_{\mathbf{k,k'}}\,.
\end{equation}
Using (\ref{eq:D_fromM}) we can then calculate the diffusion coefficient $D$ already discussed in the paper and shown in Fig.\ref{fig:D_T}.
Some details about the low temperature and high temperature limits of $D$ are given in appendix \ref{app:lowT} and in appendix \ref{app:highT}
respectively. In particular we find at low temperature
\begin{equation}
\frac{\hbar D V}{g} \sim c_1 \left(\frac{k_BT}{\rho g}\right)^4   \, \:\:\:\:\:\: {\rm for} \:\:\:\:\:\: \frac{k_BT}{\rho g} \to 0
\end{equation}
The constant $c_1=0.3036$ is calculated numerically. 

\noindent{\bf The constant term:}
We now come to the constant term ${\cal C}$ defined in (\ref{eq:calC}). By integrating formally 
$(d/dt)(t\vec{X})$ between zero and infinity and by using (\ref{eq:Xdot}), we obtain
\begin{equation}
0=\int_0^\infty dt \, \vec{X}(t) + M  \int_0^\infty dt \, t\vec{X}(t)
\end{equation}
and finally
\begin{equation}
{\cal C}=-2(P \vec{A}) \cdot M^{-2} \vec{X}(0) \,.
\label{eq:calC_final}
\end{equation}
We show in Fig.\ref{fig:c0_T} the constant ${\cal C}$ obtained from (\ref{eq:calC_final}) as a function of temperature. At low temperature we get
\begin{equation}
\frac{{\cal C} V}{\xi^3} \sim c_2 \left( \frac{k_BT}{\rho g} \right)^{-1} \, \:\:\:\:\:\:{\rm for} \:\:\:\:\:\: \frac{k_BT}{\rho g} \to 0
\end{equation}
The constant $c_2=-0.2033$ is calculated numerically. 
Note that, contrarily to the coefficients $\mathcal{A}$ and
$\mathcal{B}$, the coefficient $\mathcal{C}$ does not tend to
zero for $T\to 0$, on the contrary it diverges.
However, the typical decay time $\tau_c$ of the correlation function
$C(t)$ also
diverges in this limit, as we shall see in what follows.

\noindent{\bf The correlation function $C(t)$:}
The phase derivative correlation function $C(t)$ was defined
in (\ref{eq:C(t)}).
Restricting for simplicity to the system being prepared in the
microcanonical ensemble (as we have seen,
in the general case, $C(t)$ deviates
from the microcanonical value $C_{\rm mc}(t)$ by an additive constant),
we show in Fig.\ref{fig:varphi}-Top  
the function $C_{\rm mc}(t)$ in the low temperature case 
$k_BT/\rho g=0.2$. $C_{\rm mc}(t)$ is obtained by (\ref{eq:C_x(t)}), 
integrating equation (\ref{eq:xdot}) in time by Euler's method. 
Correspondingly, we calculate the variance of the condensate
accumulated phase 
as a function of time from (\ref{eq:varphi}) 
and we compare it to its asymptotic behavior (\ref{eq:genVarphi}).
On the same figure, see Fig.\ref{fig:varphi}-Bottom,
we show $C_{\rm mc}(t)$ in log-log scale to point out significant deviations from the exponential behavior: 
$C(t)$ rather decays as a power law;
the Gaussian model of \cite{PRA_Micro} 
at large times gives $C_{\rm mc}(t)\propto 1/t^3$ which we also plot in 
the figure for comparison. 

\begin{figure}[htb]
\centerline{\includegraphics[width=8.5cm,clip=]{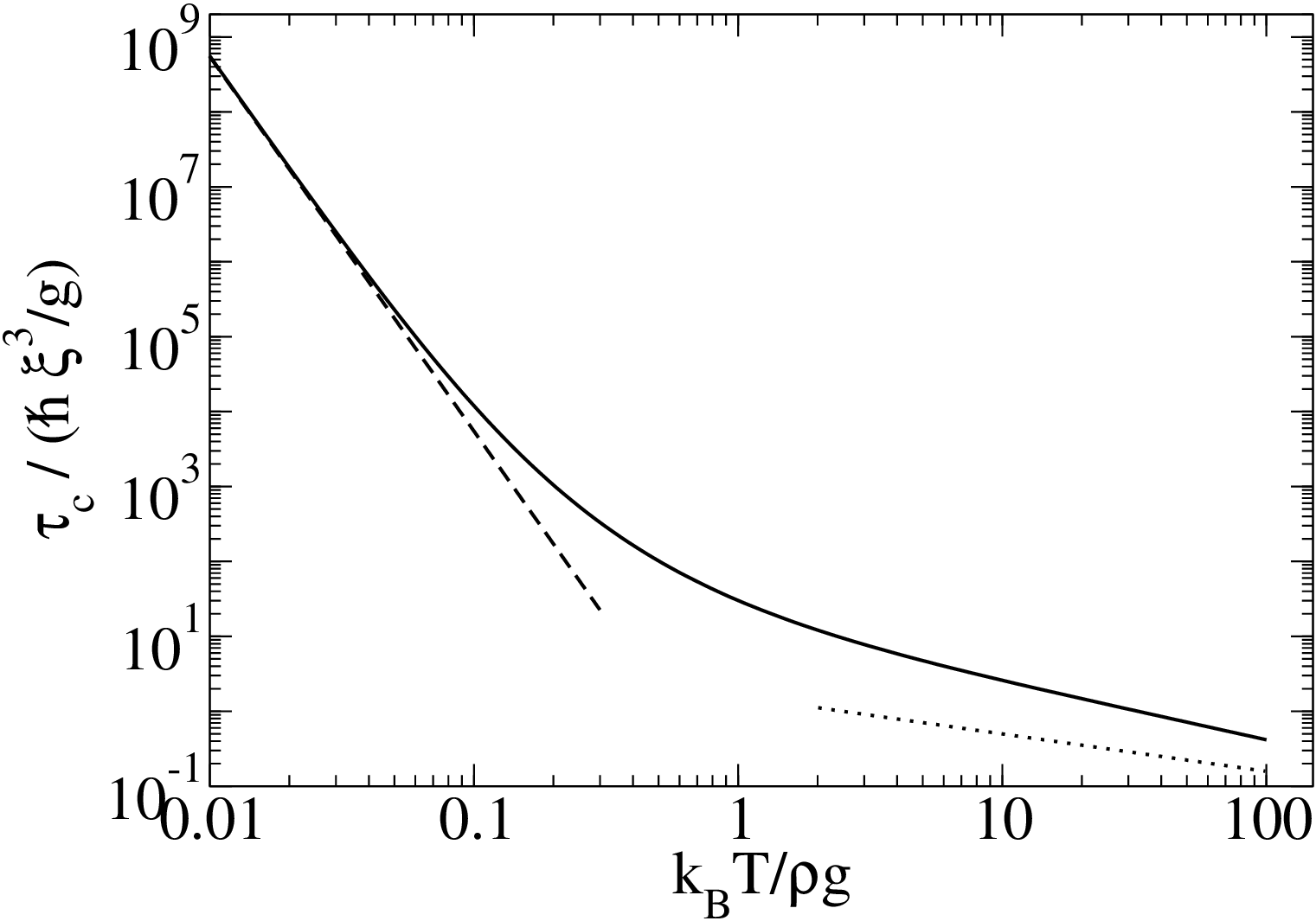}}
\caption{Typical decay time of the correlation function $C(t)$, after which the asymptotic behavior of the phase is observed. Full line: 
numerical result from the solution of (\ref{eq:tauc}). Dashed line: analytical prediction at low temperature: $y=0.05472 x^{-5}$
(see Appendix \ref{app:lowT}). Dotted line: naive prediction 
for the high temperature scaling:
$y \propto x^{-1/2}$ (see Appendix \ref{app:highT}).
The healing length $\xi$ is such that $\hbar^2/2m\xi^2 = \rho g$.
\label{fig:tca_T}}
\end{figure}

\noindent{\bf Characteristic time to reach the asymptotic regime:}
The asymptotic regime for the phase variance is reached after a transient that is the typical decay time of the correlation function $C(t)$.
An estimation of this time is 
\begin{equation}
\tau_c \equiv \frac{D}{C_{\rm mc}(0)} \,. \label{eq:tauc}
\end{equation}
This is only an estimation  since, as we have seen,
$C_{\rm mc}(t)$ is not an exponential function $\propto \exp(-t/\tau_c)$. 
A plot of $\tau_c$ as a function of temperature is shown in Fig.\ref{fig:tca_T}.
At low temperature 
\begin{equation}
\frac{g \tau_c}{\hbar \xi^3} \sim c_3 \left( \frac{k_BT}{\rho g} \right)^{-5} \, \:\:\:\:\:\:{\rm for} \:\:\:\:\:\: \frac{k_BT}{\rho g} \to 0
\end{equation}
The constant $c_3=0.05472$ is calculated numerically. 

In table \ref{tab:values} we give the numerical values of the relevant parameters for 10 reduced temperatures in the range  $0.1-100$.
\begin{table}[htdp]
\caption{Numerical values of the relevant quantities. $\xi$ is the healing length: $\hbar^2/2m\xi^2=\rho g$. ${\cal A}$ is given for energy fluctuations of the canonical ensemble, and $C(0)$ for the microcanonical ensemble. } 
\begin{center}
\begin{tabular}{|c|c|c|c|c|c|}
\hline
$\frac{k_BT}{\rho g}$ & $\frac{\hbar D V}{g}$       & $\frac{{\cal C}V}{\xi^3}$ & $\frac{C_{\rm mc}(0) V \hbar^2 \xi^3}{g^2}$ & $\frac{g \tau_c}{\hbar \xi^3}$ 
	& $\frac{ {\cal A}_{\rm can}^{1/2}\hbar V^{1/2} }{(\rho g) a \xi^{1/2}}$ \\
\hline
$0.1$     &  $2.130\times 10^{-5}$ &  $-2.227$     &  $1.784\times 10^{-9}$ &   $11940$   &  0.02397\\
\hline
$0.2$     &  $2.142\times 10^{-4}$ &  $-1.426$     &  $2.046\times 10^{-7}$ &   $1046$      & 0.1092 \\
\hline
$0.5$     &  $3.163\times 10^{-3}$ &  $-1.286$     &  $3.105\times 10^{-5}$  &   $101.9$    & 0.6037 \\
\hline
$1$        &  $1.911\times 10^{-2}$ &  $-1.726$     &  $6.337\times 10^{-4}$  &   $30.16$    & 1.7557 \\
\hline
$2$        &  $9.626\times 10^{-2}$ &  $-2.886$     &  $7.939\times 10^{-5}$  &   $12.12$    & 4.3682 \\
\hline
$5$        &  $0.638$                       &  $-6.691$     &  $0.134$                        &   $4.746$    & 12.276 \\
\hline
$10$      &  $2.280$                       &  $-12.95$     &  $0.880$                        &   $2.590$    & 24.542 \\
\hline
$20$      &  $7.323$                       &  $-24.48$     &   $4.971$                       &   $1.473$    & 46.598 \\
\hline
$50$      &  $30.14$                       &  $-53.35$     &   $42.06$                       &   $0.716$     & 103.10\\
\hline
$100$    &  $81.60$                       &  $-91.59$     &   $195.8$                       &   $0.417$    & 182.94 \\
\hline
\end{tabular}
\end{center}
\label{tab:values}
\end{table}%

\section{Influence of particle losses on the super-diffusive phase spreading}
\label{sec:losses}
For an isolated system with energy fluctuations in the initial state, we have seen that the correlation function $C(t)$ of the condensate phase derivative does not vanish at long times and the condensate phase spreading is super diffusive. In presence of particle losses, unavoidable in real experiments, the system is not isolated and the total energy is not conserved so that one may wonder whether the super diffusive term is still present. We show in this section that this is indeed the case, in a regime where the {\it fraction} of particles 
lost during the decay time $\tau_c$ of the correlation function $C(t)$ is small, a condition satisfied in typical experimental conditions. 

\begin{figure}[htb]
\centerline{\includegraphics[width=8.5cm,clip=]{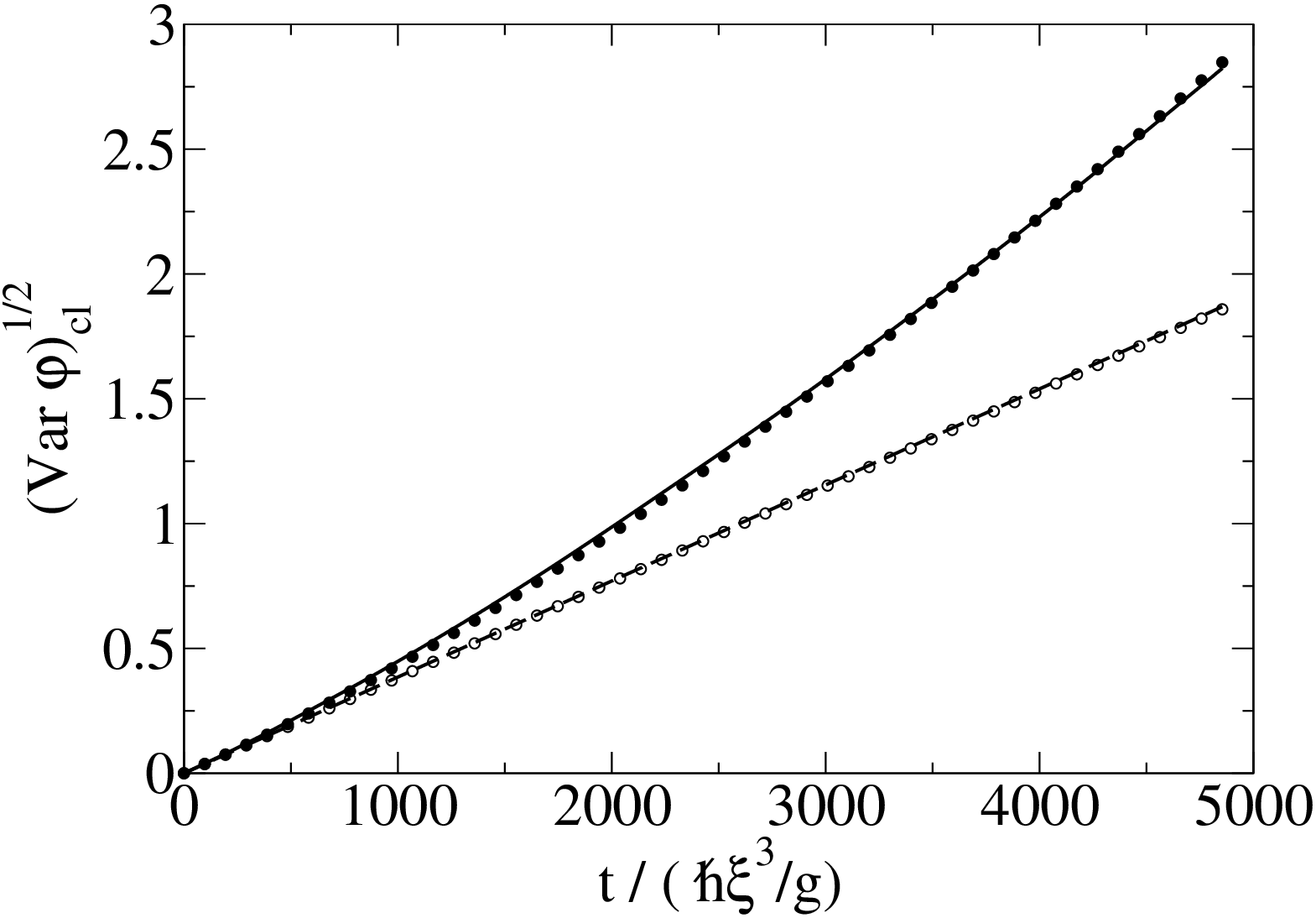}}
\caption{Condensate accumulated phase standard deviation 
as a function of time with and without one body losses in a classical field model. Solid line: simulation
for $\Gamma=1.555\times 10^{-5} g/(\hbar\xi^3)$. 
Dashed line: simulation without losses. Black discs: lossy ergodic model (see text) for $\Gamma=1.555\times 10^{-5} g/(\hbar\xi^3)$.
Circles: prediction of the ergodic theory (no losses). The initial atom number is $N(0)=4\times10^5$, $\rho(0) g=1798.47  \hbar^2/(m V^{2/3})$, $k_BT/\rho(0) g=2.95$.
A spatial box of sizes $L_1$, $L_2$, $L_3$ and volume $V=L_1 L_2 L_3$ is used with periodic boundary conditions. The squared box sizes are in the ratio
$\sqrt{2}:(1+\sqrt{5})/2:\sqrt{3}$. Note that here, contrarily to previous figures, the variance is directly given and was not
divided by the factor $\xi^3/V$ (here $\xi^3/V\simeq 4.64\times 10^{-6}$). 
For a typical atomic density of $\rho(0)=1.2 \times 10^{20}$ atoms/m${}^3$, taking the ${}^{87}$Rb mass and scattering length
$a=5.3$ nm, our parameters correspond to $1/\Gamma=20$s, $\rho(0) g/(2\pi \hbar) \simeq 950$ Hz, $T\simeq 0.14 \mu$K or $T \simeq 0.3 T_c$,
and the temporal unit $\hbar\xi^3/g\simeq 0.31$ms. A number of 1200 realizations is used in each simulation, and the energy
cut-off corresponds to a maximal Bogoliubov eigenenergy equal to $k_B T$.
The variance of the total energy in the initial state is $1.5 \times 10^{11} \hbar^4/(m V^{2/3})^2$, resulting from sampling the canonical ensemble in the Bogoliubov approximation. This value is larger than the one predicted by the Bogoliubov theory by a factor 1.3 due to non negligible interactions among the Bogoliubov modes.  A lossless relaxation phase of a duration $500 \hbar\xi^3/g$ is performed after the sampling. 
\label{fig:loss_varphi}}
\end{figure}

We first perform a classical field simulation with one body losses of rate constant $\Gamma$:  during the infinitesimal time interval $dt$, a quantum jump may occur with a probability $\Gamma N dt$ where $N$ is the number of particles just before the jump. If the jump occurs, a particle is lost which corresponds in the classical field model
to a renormalization of the field $\psi({\bf r}) \to [(N-1)/N]^{1/2} \psi({\bf r})$. In between jumps the field evolves with the usual non-linear Schr\"odinger equation:
\begin{equation}
i\hbar \partial_t \psi = - \frac{\hbar^2}{2m} \Delta\psi + g|\psi|^2\psi.
\end{equation}
This results from the interpretation of the classical field in terms of an Hartree-Fock Ansatz for the quantum system state as detailed in Appendices
\ref{app:modele} and \ref{app:modeleQ}.
The result for the condensate accumulated 
phase standard deviation as a function of time is shown in Fig. \ref{fig:loss_varphi}, in the absence (dashed line) and in presence (solid line) of losses. It is apparent that, for the parameters taken in this figure, the spreading of the phase up to a standard deviation of order unity
is only weakly affected by the particle losses. We also find that the phase spreading is in fact accelerated by the losses and becomes effectively super-ballistic.
As we now show,  this is due to the fact that the losses introduce 
particle number fluctuations that grow in time. 

In order to understand the numerical results we use 
a heuristic extension of the ergodic model in presence of 
losses. In the model there are two dynamical variables: the total energy and 
the total number of particles. We assume that in between two loss events the
condensate phase evolves according to
\begin{equation} 
\dot{{\theta}}(t)=-\frac{\mu_{\rm mc}(E,N)}{\hbar}
\label{eq:thdot_mod}
\end{equation} 
where $\mu_{\rm mc}$ is the chemical potential in the microcanonical ensemble 
of energy $E$ for a system with $N$ particles.
When a loss event occurs, $N$ is obviously changed into $N-1$.
For the energy change one has to consider separately the kinetic and the 
interaction energies: $E_{\rm kin}$ is a quadratic function of $\psi$ and
is changed into $[(N-1)/N] E_{\rm  kin}$. The interaction energy is a quartic
function of $\psi$ and is changed into $[(N-1)/N]^2 E_{\rm  int}$. 
When a jump occurs we then take 
\begin{eqnarray}
E'&=& \frac{N-1}{N} \, \langle E_{\rm kin} \rangle_{\rm mc} (E,N) \nonumber \\
&+& \left( \frac{N-1}{N} \right)^2 \langle E_{\rm int} \rangle_{\rm mc} (E,N) 
\label{eq:E'} \\
N'&=&N-1
\label{eq:N'}
\end{eqnarray}
where the prime indicates the quantities after the jump and where
$\langle E_{\rm kin} \rangle_{\rm mc}$ and 
$\langle E_{\rm int} \rangle_{\rm mc}$ are the mean kinetic and interaction 
energies in the microcanonical ensemble 
\cite{note_cin_int}.
To calculate the microcanonical averages and the chemical potential, we
rely on Bogoliubov theory. In the classical field model:
\begin{eqnarray}
\label{eq:itcfm1}
\mu_{\rm mc}(E,N)&=& \frac{gN}{V} + \frac{g}{V} \frac{E-E_0}{\cal M}
\sum_{{\bf k}\ne {\bf 0}}  \left( \frac{\hbar^2 k^2}{2m} + \frac{2gN}{V}\right)^{-1}  \\
\label{eq:itcfm2}
\langle E_{\rm int} \rangle_{\rm mc}(E,N) &=& E_0
+\frac{gN}{V} \frac{E-E_0}{\cal M}
\sum_{{\bf k}\ne {\bf 0}} \left( \frac{\hbar^2 k^2}{2m} + \frac{2gN}{V}\right)^{-1} \\
\label{eq:itcfm3}
\langle E_{\rm kin} \rangle_{\rm mc}(E,N) &=&
 E- \langle E_{\rm int} \rangle_{\rm mc}(E,N) \,,
\end{eqnarray}
where ${\cal M}$ is the number of Bogoliubov modes and $E_0=gN^2/2V$ is the 
ground state energy. 

We have performed a Monte Carlo simulation of this model. 
The initial energy is obtained sampling a Gaussian distribution with a mean
energy given by Bogoliubov theory and with the same variance as in 
the classical field simulations (see caption of Fig.\ref{fig:loss_varphi}). 
The results for the condensate phase variance (symbols) are compared 
with the classical field simulation with and without losses in 
Fig.\ref{fig:loss_varphi}. A good agreement is found.

To go further, we analytically solve this model to first order in the loss
rate constant $\Gamma$.
As detailed in appendix \ref{app:modele}, we obtain the simple result:
\begin{multline}
{\mbox{Var}} \, \hbar\varphi (t) \simeq ({\mbox{Var}}\, \mu)  t^2 
+\left( \langle \mu \, \delta\mu\rangle -
\langle\mu\rangle \langle\delta\mu\rangle \right)
\Gamma N t^3 \\
+\frac{1}{3} \langle \delta\mu^2\rangle  \Gamma N t^3.
\label{eq:raF}
\end{multline}
Here $N$ is the initial atom number,
$\mu$ is the initial chemical potential, and $\delta \mu=\mu'-\mu$ is
its change after the first loss event.
After an explicit calculation, in the limit $k_BT \gg \rho g$, this reduces to
\begin{multline}
{\mbox{Var}} \, \varphi (t) \simeq \left( \frac{d\mu_{\rm mc}}{dE}\right)^2(\bar{E},N) {\rm{Var}}\, E \; \frac{t^2}{\hbar^2} \\
+ \frac{1}{3}\Gamma Nt^3 \left( \frac{g}{\hbar V}\right)^2 \\
+ \Gamma Nt^3 \left( \frac{g}{\hbar V}\right)^2 O\left( \frac{\langle \delta N \rangle}{N} \right) \,,
\label{eq:rap}
\end{multline}
where $\langle \delta N \rangle/N \ll 1$ is the fraction of non condensed particles.
The first term in the right hand side of (\ref{eq:rap}) is the classical field version of the result (\ref{eqs:varphi}) without losses. The third term is negligible
as compared to the second one in the present regime of a small non condensed fraction.
The second term, independent of the temperature, is the result that one would have at zero temperature in presence of losses at short times
$(\Gamma t \ll 1)$. This term has a simple physical interpretation: in presence of fluctuations in the initial number of particles for a lossless pure condensate, 
the condensate accumulated phase grows quadratically in time with a variance $(d\mu/dN)^2 t^2 \mbox{Var}\,N/\hbar^2$, where $\mu=gN/V$. 
For a lossy pure condensate with initially exactly $N$ particles, $\mbox{Var}\,N\simeq \Gamma Nt$ 
so that one indeed expects 
${\mbox{Var}} \, \varphi (t) \propto (g/\hbar V)^2 \Gamma N t^3$.
Actually, at $T=0$ 
it is possible to calculate exactly ${\mbox{Var}} \, \varphi (t)$ in presence of losses (see appendix \ref{app:modele}):
\begin{equation}
\left[ {\mbox{Var}} \, \varphi (t) \right]_{T=0} = \left( \frac{g}{V\hbar \Gamma}\right)^2 N \left[ 1 - 2\Gamma t \, e^{-\Gamma t} - e^{-2\Gamma t}\right] \,.
\label{eq:casT=0}
\end{equation}
This zero temperature result even extends to the quantum case for a pure
condensate, see Appendix \ref{app:modeleQ}, so that one may hope
that the form of the classical field result (\ref{eq:rap}) 
extends to the quantum reality. 
To be complete we also give the exact value of the correlation function $\langle \hat{a}_0^\dagger(t) \hat{a}_0(0) \rangle$ in the quantum case for a pure 
condensate with initially $N$ particles and subject to one-body losses:
\begin{equation}
\langle \hat{a}_0^\dagger(t) \hat{a}_0(0) \rangle_{T=0}=e^{-\Gamma t/2} N \left[ e^{-\lambda t}+ \frac{\Gamma}{\lambda}
\left( 1-e^{-\lambda t} \right)\right]^{N-1} \label{eq:result}
\end{equation}
with $\lambda=\Gamma-\frac{ig}{\hbar V}$. This can be obtained by applying the quantum regression theorem using (\ref{eq:ME}) expressed in the
Fock basis. The same result (\ref{eq:result}) can be obtained using the exact result for a two mode model Eq.(125) of \cite{EPJB}, and assuming 
that the second mode of infinitesimal population experiences no interactions and no particle losses.

\section{Conclusions}
\label{sec:conclusions}
In conclusion we have presented a full quantitative quantum solution to the long standing problem of the decoherence of a condensate due to its interactions
with quasi-particles in the non-condensed modes: the growth of the variance
of the condensate accumulated phase involves in general {\it both} a quadratic term and a linear term in time, with coefficients that we have determined within a {\it single} theoretical frame, quantum kinetics. As we have discussed, our findings may be directly tested with state-of-the-art technology, and they may stimulate systematic experimental investigation of this problem, both fundamental and crucial for future applications of condensates in matter wave interferometry. 
\section{acknowledgments}
We thank Carlos Lobo for a careful reading of the manuscript and we acknowledge financial support from the PAN/CNRS collaboration. We thank J. Est\`eve, J. Reichel, F. Gerbier and J. Dalibard for useful discussions. 

\appendix

\section{Equations for $x_{{\bf q}}(t)$}
\label{sec:eqs_detail}
We detail here the derivation of equation (\ref{eq:xdot}) for the correlation functions $x_{\bf q}(t)$. We assume
that (i) the density matrix $\hat{\rho}$ of the gas is a statistical mixture of eigenstates of the Bogoliubov Hamiltonian $\hat{H}_{\rm Bog}$ given by (\ref{eq:Hbog})
\begin{equation}
\hat{\rho}=
\sum_{\{ n_{\bf k} \}} {\cal P}(\{ n_{\bf k} \}) |\{ n_{\bf k} \}\rangle \langle\{ n_{\bf k} \}| \,,
\label{eq:rho}
\end{equation}
and (ii) for a given initial Fock state $|\{ n_{\bf k} \}\rangle$, the evolution of the expectation values 
$n_{{\bf q}}(t)=\langle \{ n_{\bf k}(0) \}|\hat{n}_{\bf q}(t)|\{ n_{\bf k}(0) \}\rangle$
are given by the kinetic equations (\ref{eq:kin}). We then have
\begin{eqnarray}
\langle \hat{n}_{\bf q_1}(t) \hat{n}_{\bf q_2}(0)\rangle &=& \sum_{\{ n_{\bf k}(0) \}}  {\cal P}(\{ n_{\bf k}(0) \}) n_{\bf q_2}(0) \nonumber \\
&\times& \langle \{ n_{\bf k}(0) \}|\hat{n}_{\bf q_1}(t)|\{ n_{\bf k}(0) \}\rangle \,,
\end{eqnarray}
and 
\begin{eqnarray}
\frac{d}{dt}\langle \delta \hat{n}_{\bf q_1}(t) \delta \hat{n}_{\bf q_2}(0)\rangle &=& \sum_{\{ n_{\bf k}(0) \}}  {\cal P}(\{ n_{\bf k}(0) \}) \delta n_{\bf q_2}(0) 
 \nonumber \\
&\times&  \sum_{\bf q'} M_{\bf q_1,q'}  \delta n_{{\bf q'}}(t) \,,
\label{eq:amul}
\end{eqnarray}
where the matrix $M$ is obtained by linearization of equations (\ref{eq:xdot}).
We have introduced
\begin{equation}
\delta {n}_{\bf q}(t) = 
\langle \{ n_{\bf k}(0) \}|\hat{n}_{\bf q}(t)|\{ n_{\bf k}(0) \}\rangle
-
\langle \hat{n}_q\rangle
\end{equation}
where we recall that $\langle\ldots\rangle$ is the expectation value
in the state of the system.
By multiplying (\ref{eq:amul}) by $A_{\bf q_2}$, summing over ${\bf q_2}$,
and approximating $\langle \hat{n}_q\rangle$ with
$\bar{n}_q$ of (\ref{eq:nstat}), which is justified in the present regime of
large system size and 
weak relative energy fluctuations,  we obtain 
(\ref{eq:xdot}).

Using the rotational invariance of $x_{{\bf k}}$ as a function of ${\bf k}$ and
the delta of conservation of energy we can explicitly integrate over the angular variables and we obtain the simple integral equations
that we now detail.
We introduce dimensionless quantities $\check{Q}$. 
Momenta are rescaled by the inverse of the healing length $\xi=(\hbar^2/2m\rho g)^{1/2}$, energies
are rescaled by the Gross-Pitaevskii chemical potential $\rho g$, and rates are expressed in units of $g/(2\pi^2 \xi^3 \hbar)$: 
\begin{eqnarray}
 \check{q}&=& q \left(\frac{\hbar^2}{2m\rho g}\right)^{1/2} = q \xi,\\
\check{\epsilon}_{q} &=&\frac{\epsilon_q}{\rho g}  = [\check{q}^2(\check{q}^2+2)]^{1/2} \\
\check{\Gamma}_{q} &=&  \frac{2\pi^2 \xi^3 \hbar}{g} \Gamma_q\,.
\end{eqnarray}
As a consequence, the mean occupation number $\bar{n}_q$ is a function of $\check{q}$ and of the ratio
$k_BT/\rho g$ only, and the mode amplitudes $U_q, V_q$ are functions of $\check{q}$ only. 
Expressing the time in reduced units, we then have
\begin{equation}
\dot{x}_{q}(t)=-\check{\Gamma}_{q} x_{q}(t) + \check{I} \,.
\label{eq:dotx_int}
\end{equation}
The integral $\check{I}$ is:
\begin{eqnarray}
\frac{\check{I}}{2\pi} &=& \int_0^\infty d\check{k}  \left( {\cal A}_{k,q}^{k'} \right)^2  \frac{\check{k} (\check{\epsilon}_k+\check{\epsilon}_q)
(\bar{n}_{k'}-\bar{n}_q)}{\check{q}(\check{k}'^2+1)}  x_k(t) \nonumber \\
 &+& \int_q^\infty d\check{k}   \left({\cal A}_{k'',q}^{k} \right)^2  \frac{\check{k} (\check{\epsilon}_k-\check{\epsilon}_q)
 (1+\bar{n}_{k''}+\bar{n}_q)}{\check{q}(\check{k}''^2+1)} x_k(t) \nonumber \\
 &+& \int_0^q d\check{k}  \left({\cal A}_{k,k''}^{q} \right)^2 \frac{\check{k} (\check{\epsilon}_q-\check{\epsilon}_k)
 (\bar{n}_{k''}-\bar{n}_q) }{\check{q}(\check{k}''^2+1)} x_k(t) \,,
\label{eq:final_int} 
\end{eqnarray}
with
\begin{eqnarray}
\check{k}'^2&=& \sqrt{1+(\check{\epsilon}_k+\check{\epsilon}_q)^2}-1 \\
\check{k}''^2&=& \sqrt{1+(\check{\epsilon}_k-\check{\epsilon}_q)^2}-1 \,. 
\end{eqnarray}
The damping rate $\Gamma_q$ is the sum of the Beliaev and Landau damping rates already given in \cite{PRA_Micro}:
\begin{equation}
\check{\Gamma}_{q}=\check{\Gamma}_{q}^L+\check{\Gamma}_{q}^B
\end{equation}
with
 \begin{equation}
\frac{\check{\Gamma}_q^L}{2\pi}= 
\int_0^{+\infty} \!\! d\check{k} \,  \, \left(  {\cal A}_{k,q}^{k'} \right)^2 
 \frac{\check{k}(\check{\epsilon}_k+\check{\epsilon}_q) (\bar{n}_k-\bar{n}_{k'})}
{\check{q}(\check{k}'^2+1)}
\label{eq:GammaL}
\end{equation}
and
\begin{equation}
\frac{\check{\Gamma}_q^B}{\pi} =  \int_0^{\check{q}} d\check{k}\, \left(  {\cal A}_{k,k''}^{q} \right)^2
\frac{\check{k}(\check{\epsilon}_q-\check{\epsilon}_k)(1+\bar{n}_k+\bar{n}_{k''})}{\check{q}(\check{k}''^2+1)} \,.
\label{eq:GammaB}
\end{equation}
Introducing 
\begin{eqnarray}
\check{M}&=& \frac{2\pi^2 \xi^3 \hbar}{g} M \\
\check{\vec{A}}&=& \frac{\hbar V}{g} \vec{A} \\
\check{\vec{X}}(t)&=& \frac{\hbar V}{g} \vec{X}(t)
\end{eqnarray}
one has
\begin{eqnarray}
\frac{\hbar D V}{g}&=&- \int_0^\infty \check{k}^2 d\check{k} (P\check{\vec{A}})_{k} (\check{M}^{-1} \check{\vec{X}}(0))_{k} \\
\frac{{\cal C}V}{2\pi^2 \xi^3} &=&-2 \int_0^\infty \check{k}^2 d\check{k} (P\check{\vec{A}})_{k} (\check{M}^{-2}  \check{\vec{X}}(0))_{k} \\
C_{\rm mc}(t) &=& \frac{g^2}{2\pi^2 V \hbar^2 \xi^3}\int_0^\infty \check{k}^2 d\check{k} (P\check{\vec{A}})_{k} \check{X}(t)_{k} \,.
\end{eqnarray}

\section{Case of a classical field}
\label{app:class}

We consider a discrete model for a classical field $\psi({\bf r})$ in
three dimensions. The lattice spacing is $l$ 
along the three directions of space. 
We enclose the field in a spatial box of volume $V$ with periodic boundary conditions. 
Then the field can be expanded over the plane waves
\begin{equation}
\psi({\bf r})=\sum_{\bf k} a_{\bf k} 
\frac{e^{ \, i \, {\bf k}\cdot{\bf r}}}{\sqrt{V}} \,,
\label{eq:planewaves}
\end{equation}
where $\mathbf{k}$ is restricted to the first Brillouin zone,
$\mathbf{k}
\in {\cal D}\equiv [-\pi/l,\pi/l[^3$.
The lattice spacing corresponds to an energy cut-off such that the highest Bogoliubov energy on the lattice is $\epsilon_{k_{\rm max}}=k_BT$.

The classical limit in the kinetic equations is obtained by taking:
$\bar{n}_k+1\simeq \bar{n}_k \to \bar{n}^{\rm cl}_k=k_B T/\epsilon_k$ in the equation (\ref{eq:dotx_int})
for $x_{\bf q}$. 
In the units already introduced in Appendix \ref{sec:eqs_detail} one then has:
\begin{equation}
\dot{x}_{\bf q}(t) = - \check{\Gamma}^{\rm cl}_{\bf q} \, x_{\bf q}(t) + \check{I}_{\rm cl}.
\end{equation}
We have introduced
\begin{eqnarray}
\frac{\check{I}_{\rm cl}}{2} & = & \int_{\cal \check D} d^3\check{k} \left( \mathcal{A}_{k, k'}^q \right)^2
(\bar{n}^{\rm cl}_{k'} - \bar{n}^{\rm cl}_q) \delta(\check{\epsilon}_k + \check{\epsilon}_{k'}
-\check{\epsilon}_q)  x_{\bf k}(t) \nonumber \\
{}&+& \int_{\cal \check D} d^3\check{k} \left( \mathcal{A}_{k, q}^{k''} \right)^2
(\bar{n}^{\rm cl}_{k''}-\bar{n}^{\rm cl}_q) \delta(\check{\epsilon}_k+\check{\epsilon}_q 
-\check{\epsilon}_{k''}) x_{\bf k}(t) \nonumber \\
{}&+& \int_{\cal \check D} d^3\check{k}\left( \mathcal{A}_{k', q}^{k} \right)^2
(\bar{n}^{\rm cl}_{k'} + \bar{n}^{\rm cl}_q) \delta(\check{\epsilon}_{k'}+\check{\epsilon}_q 
-\check{\epsilon}_k)  x_{\bf k}(t) \, . \nonumber\\
\end{eqnarray}
The integrals are restricted to the domain 
${\cal \check D}=[-\pi\xi/l,\pi\xi/l[^3$ and 
\begin{eqnarray}
{{\bf k}'}&=&{\bf q} - {\bf k} + \frac{2\pi}{l}{\bf n} \, ,\:\:\:\:\:\: {\bf n} \in {\mathbb{Z}}^3 \\
{{\bf k}''}&=&{\bf q} + {\bf k} + \frac{2\pi}{l}{\bf m} \, ,\:\:\:\:\:\: {\bf m} \in {\mathbb{Z}}^3  ,
\end{eqnarray}
where ${\bf m}$ and ${\bf n}$ are such that ${\bf k'},{\bf k''}\in {\cal D}$.
Indeed the presence of the lattice implies the existence of unphysical Umklapp processes, such that 
${\bf n}\neq\mathbf{0}$ or ${\bf m}\neq\mathbf{0}$ 
(see \cite{PRA_Micro}), that we include in the 
classical kinetic theory. 

The damping rate in the classical field model
$\check{\Gamma}^{\rm cl}_{\bf q}$ is the sum of Beliaev and Landau damping
rates $\check{\Gamma}^{\rm cl}_{\bf q}=
\check{\Gamma}^{{\rm cl},B}_{\bf q}+\check{\Gamma}^{{\rm cl},L}_{\bf q}$ with:
\begin{equation}
\check{\Gamma}^{{\rm cl},B}_{\bf q} = \int_{\cal \check D} d^3\check{k}  \left( \mathcal{A}_{k, k'}^q \right)^2 (\bar{n}^{\rm cl}_k + \bar{n}^{\rm cl}_{k'}) 
\delta(\check{\epsilon}_k + \check{\epsilon}_{k'}
-\check{\epsilon}_q) \, ,
\end{equation}
\begin{equation}
\check{\Gamma}^{{\rm cl},L}_{\bf q} = 2\, \int_{\cal\check  D} d^3\check{k} \left( \mathcal{A}_{k, q}^{k''} \right)^2 (\bar{n}^{\rm cl}_k - \bar{n}^{\rm cl}_{k''}) 
\delta(\check{\epsilon}_k + \check{\epsilon}_q-\check{\epsilon}_{k''}) \, . 
\end{equation}

From the kinetic equations in the classical model,
 $\dot{\vec{x}}(t)=M_{\rm cl} \vec{x}(t)$, one 
has the classical diffusion coefficient in the form:
\begin{equation}
\frac{\hbar D_{\rm cl} V}{g}=-\int_{\cal\check  D} d^3\check{k}\,  (P \check{\vec{A}})_{\bf k}\, (\check{M}_{\rm cl}^{-1} 
\check{\vec{X}}(0))_{\bf k} \, .
\end{equation}
Paradoxically the lattice with the relatively low energy cut-off breaks the spherical symmetry of the problem making the numerical solution heavier
than in the quantum case. 

The classical field simulations were performed as in \cite{PRA_Micro} on a lattice with a few percent anisotropy,
except for the free dispersion relation of the matter wave on the grid:
here the usual parabolic dispersion relation $E_k = \hbar^2 k^2/2m$
was used.

\section{State of the system and quantum averages}
\label{app:quantumave}
In this appendix we establish some useful relations among different averages. In particular we wish to express the expectation value
of $\hat{O}$ defined in equation (\ref{eq:def_ave}) in terms of canonical averages where the temperature $T$ is chosen such that 
$ \langle \hat{H}_{\rm Bog} \rangle_{\rm can}(T)=\langle \hat{H}_{\rm Bog} \rangle \equiv \bar{E}$. 
First of all we expand the function $\langle \hat{O} \rangle_{\rm mc}(E)$ around its value for the average energy:
\begin{eqnarray}
\langle \hat{O} \rangle_{\rm mc}(E)&=&\langle \hat{O} \rangle_{\rm mc}( \bar{E} ) + (E-\bar{E})\frac{d \langle \hat{O} \rangle_{\rm mc}}{dE}(  \bar{E}) \nonumber \\
&+& \frac{1}{2} (E- \bar{E})^2 
\frac{d^2 \langle \hat{O} \rangle_{\rm mc}}{dE^2}(\bar{E}) + \ldots
\label{eq:expansion_O_mc}
\end{eqnarray}
We then take the average of (\ref{eq:expansion_O_mc}) over the energy distribution $P(E)$ and obtain
\begin{equation}
\langle \hat{O} \rangle=\langle \hat{O} \rangle_{\rm mc}( \bar{E} ) 
 + \frac{1}{2}  \frac{d^2 \langle \hat{O} \rangle_{\rm mc}}{dE^2} {\rm Var} 
( \hat{H}_{\rm Bog}) + \ldots
\label{eq:expansion_O}
\end{equation}
The coefficient in front of ${\rm Var} (\hat{H}_{\rm Bog}) $ in the second term in (\ref{eq:expansion_O}) appears in a first order correction,
it can thus be calculated to lowest order in the inverse
system size. By writing
explicitly 
\begin{equation}
 \langle \hat{O} \rangle_{\rm mc}(  \bar{E}(T) )  \simeq \langle \hat{O} \rangle_{\rm can}(T)
\end{equation}
and taking the derivative of this relation with respect to the temperature $T$, we obtain
\begin{equation}
 \frac{d \langle \hat{O} \rangle_{\rm mc}}{dE}(  \bar{E}(T) )  \simeq 
\frac{\frac{d}{dT}\langle \hat{O} \rangle_{\rm can}(T)}{ \frac{d   \bar{E}(T)}{dT}  } \,, 
\label{eq:O_can_mc}
\end{equation}
and
\begin{equation}
 \frac{d^2 \langle \hat{O} \rangle_{\rm mc}}{dE^2}(  \bar{E}(T) )  \simeq 
\frac{1}{\frac{d   \bar{E}(T)}{dT} } \frac{d}{dT} \left( \frac{\frac{d}{dT}\langle \hat{O} \rangle_{\rm can}(T)}{\frac{d   \bar{E}(T)}{dT} } \right)\,.
\end{equation}
On the other hand we know that
\begin{eqnarray}
\frac{d\bar{n}_k}{dT}&=&\frac{1}{k_BT^2} \epsilon_k \bar{n}_k(1+\bar{n}_k) \label{eq:C6}\\
{\rm Var}_{\rm can}(H_{\rm Bog} )&= &\sum_{{\bf k}\ne{\bf 0}} \epsilon_k^2 \bar{n}_k(1+\bar{n}_k) \nonumber \\
&=& k_BT^2 \frac{d\bar{E} }{dT} \,. \label{eq:C7}
\end{eqnarray}
We then obtain the equation
\begin{equation}
 \langle \hat{O} \rangle \simeq \langle \hat{O} \rangle_{\rm mc}(\bar{E}) + \frac{k_BT^2}{2} \eta
 \frac{d}{dT} \left( \frac{\frac{d}{dT}\langle \hat{O} \rangle_{\rm can}(T)}{\frac{d   \bar{E}(T)}{dT} } \right) \,, \label{eq:O_Omc}
 \end{equation}
with
\begin{equation}
\eta= \frac{{\rm Var}(\hat{H}_{\rm Bog})}{{\rm Var}_{\rm can}(\hat{H}_{\rm Bog})}\,.
\end{equation}
In the particular case in which the average $\langle \hat{O} \rangle$ is taken in the canonical ensemble, $\eta=1$ and we recover 
equation (B7) of \cite{PRA_Super}. If we now eliminate the microcanonical average in (\ref{eq:O_Omc}) in favor of the canonical one, we obtain the final 
formula
\begin{equation}
 \langle \hat{O} \rangle \simeq \langle \hat{O} \rangle_{\rm can}(T) + 
 \frac{k_BT^2}{2} \frac{d}{dT} \left( \frac{\frac{d}{dT}\langle \hat{O} \rangle_{\rm can}(T)}{\frac{d   \bar{E}(T)}{dT} } \right) (\eta -1)\,.
 \label{eq:aveO_final}
\end{equation}

\section{Low temperature expansion}
\label{app:lowT}

Let us consider the limit
\begin{equation}
\frac{k_BT}{\rho g} = \varepsilon \ll 1 \,.
\end{equation}
In this case the occupation numbers $\bar{n}_q$ are exponentially small unless $\check{\epsilon}_q \lesssim \varepsilon$: indeed
\begin{equation}
\bar{n}_q=\frac{1}{e^{\beta \epsilon_q}-1} = \frac{1}{e^{\check{\epsilon}_q/\varepsilon}-1}\,.
\end{equation}
We can then restrict to low energies and low momenta where the spectrum is linear
\begin{equation}
\check{\epsilon}_q \sim \sqrt{2} \check{q} \:\:\:\:\:\:{\rm for} \:\:\:\:\:\: \check{q} \to 0 \,.
\end{equation}
We thus introduce
\begin{equation}
\tilde{q}=\frac{\check{q}}{\varepsilon} \simeq \frac{\epsilon_q}{\sqrt{2}k_BT}
\end{equation}
that is a dimensionless momentum of order unity for typical Bogoliubov mode
energies of order $k_BT$.
To obtain an expansion for $\varepsilon\ll1$, we then expand the relevant dimensionless quantities in powers of $\check{q}$ which is of order $\varepsilon$:
\begin{eqnarray}
\check{\epsilon}_q&=&\sqrt{2} \tilde{q}\, \varepsilon + \frac{\sqrt{2}}{4} \tilde{q}^3 \varepsilon^3 + O(\varepsilon^5) \\
(U_q+V_q)^2&=&  \frac{\sqrt{2}}{2} \tilde{q}\, \varepsilon -  \frac{\sqrt{2}}{8} \tilde{q}^3 \varepsilon^3 + O(\varepsilon^5) \,.
\end{eqnarray}
For a general function $F(\beta \epsilon_q)$, as for example a function of $\bar{n}_q$,
\begin{equation}
F(\check{\epsilon}_q/\varepsilon) = F(\sqrt{2}\tilde{q}) +  \frac{\sqrt{2}}{4} \tilde{q}^3 F'(\sqrt{2}\tilde{q}) \varepsilon^2 + O(\varepsilon^4) \,,
\end{equation}
and for the coefficients ${\cal A}$ in equation (\ref{eq:final_int})
\begin{eqnarray}
{\cal A}_{k,q}^{k'} &=& \frac{3}{2^{7/4}} \sqrt{\tilde{q}\tilde{k}(\tilde{q}+\tilde{k})}\, \varepsilon^{3/2} + O(\varepsilon^{5/2})\\
{\cal A}_{k,k''}^{q} &=& \frac{3}{2^{7/4}} \sqrt{\tilde{q}\tilde{k}(\tilde{q}-\tilde{k})}\, \varepsilon^{3/2}  + O(\varepsilon^{5/2}) \,.
\end{eqnarray}
On can then write the low temperature version of equations (\ref{eq:final_int}), (\ref{eq:GammaL}) and (\ref{eq:GammaB}):
\begin{eqnarray}
\check{I} &\sim& \varepsilon^5 \frac{9\pi}{4} \int_0^\infty d\tilde{k} \tilde{k}^2 (\tilde{k}+\tilde{q})^2 (\bar{n}_{k+q}-\bar{n}_q) x_k \nonumber \\
  &+& \varepsilon^5 \frac{9\pi}{4} \int_q^\infty d\tilde{k} \tilde{k}^2 (\tilde{k}-\tilde{q})^2 (\bar{n}_{k-q}+\bar{n}_q+1) x_k \nonumber \\
  &+& \varepsilon^5 \frac{9\pi}{4} \int_0^q d\tilde{k} \tilde{k}^2 (\tilde{q}-\tilde{k})^2 (\bar{n}_{q-k}-\bar{n}_q) x_k 
  \label{eq:lowt_I}\\
\check{\Gamma}_L   &\sim& \varepsilon^5 \frac{9\pi}{4} \int_0^\infty d\tilde{k} \tilde{k}^2 (\tilde{k}+\tilde{q})^2 (\bar{n}_{k}-\bar{n}_{k+q}) 
\label{eq:lowt_CL} \\
\check{\Gamma}_B   &\sim& \varepsilon^5 \frac{9\pi}{8} \int_0^q d\tilde{k} \tilde{k}^2 (\tilde{q}-\tilde{k})^2 (\bar{n}_{k}+\bar{n}_{q-k}+1)   \,,
\label{eq:lowt_GB}
\end{eqnarray}
where $\sim$ stands for mathematical equivalence in the limit $\epsilon \to 0$ ($f\sim g$ if $f/g \to 1$).
In order to obtain the scaling with $\varepsilon$ of the diffusion coefficient $D$ and of the other quantities,
we expand $P \vec{A}$ and $\vec{X}(0)$:
\begin{eqnarray}
(P\check{\vec{A}})_{q}&=& \frac{\sqrt{2}}{4} \left[ \tilde{q} {\cal R} -  \tilde{q}^3 \right] \varepsilon^3 + O(\varepsilon^5) 
\label{eq:lowt_PA}\\
\check{X}(0)_q&=&\frac{\sqrt{2}}{4} F(\sqrt{2} \tilde{q}) \left[ \tilde{q} {\cal R} -  \tilde{q}^3 \right] \varepsilon^3 + O(\varepsilon^5) 
\label{eq:lowt_X(0)}
\end{eqnarray}
with
\begin{eqnarray}
F(\beta \epsilon_q)&=& \bar{n}_q(\bar{n}_q+1) \\
{\cal R}&=&\frac{\int_0^\infty d\tilde{k} \tilde{k}^6 F(\sqrt{2} \tilde{k})}{\int_0^\infty d\tilde{k} \tilde{k}^4 F(\sqrt{2} \tilde{k})} \,.
\end{eqnarray}
We then conclude that for $\varepsilon \to 0$
\begin{eqnarray}
\frac{\hbar D V}{g}&\sim& c_1 \, \varepsilon^4 \label{eq:sca_D} \\
\frac{{\cal C}V}{\xi^3} &\sim& c_2  \, \varepsilon^{-1} \label{eq:sca_C}\\
\frac{g \tau_c}{\hbar \xi^3}&\sim& c_3 \, \varepsilon^{-5}  \\
\frac{C_{\rm mc}(0) \hbar^2V \xi^3}{g^2}&\sim& c_4 \, \varepsilon^9 \label{eq:sca_C(0)}\,.
\end{eqnarray}
In (\ref{eq:sca_D})-(\ref{eq:sca_C(0)}) a factor $\varepsilon^3$ comes from $d\tilde{k} \tilde{k}^2$ in the Jacobian.
The numerical coefficients $c_1$ to $c_4$ can be calculated numerically using the expanded expressions (\ref{eq:lowt_I})-(\ref{eq:lowt_X(0)}), or using the original expressions and extrapolating the result for $k_BT/\rho g\to 0$.

\section{High temperature}
\label{app:highT}
Let us now consider the high temperature limit 
\begin{equation}
\frac{k_BT}{\rho g} \gg 1 \,.
\end{equation}
A naive approach then consists in replacing the dispersion relation of quasiparticles by the free particle one
\begin{equation}
\check{\epsilon}_q=\sqrt{\check{q}^2(\check{q}^2+2)}\simeq \check{q}^2  \:\:\:\:\:\:{\rm for} \:\:\:\:\:\: \check{q} \to \infty \,,
\end{equation} 
and introduce the rescaled dimensionless momentum 
\begin{equation}
\tilde{\tilde{q}}=\frac{\check{q}}{\sqrt{\frac{k_BT}{\rho g} }}
\end{equation}
so that $\tilde{\tilde{q}}^2\simeq \epsilon_q/k_BT.$
In this limit $U_q\to 1$, $V_q\to 0$, ${\cal A}_{k,k'}^q\to 1$, and
\begin{equation}
\bar{n}_q=\frac{1}{e^{\beta \epsilon_q}-1} \to  \frac{1}{ e^{\tilde{\tilde{q}}^2}-1}\,.
\end{equation}
To lowest order, the integral $\check{I}$ and the rate $\check{\Gamma}$ are 
$\propto (k_BT/\rho g)^{1/2}$ while
$\check{\vec{X}}(0)$ and $P \check{\vec{A}}$ do not depend on $k_B T/\rho g$.
Similarly to the low temperature limit 
one could then deduce the high temperature scaling of the relevant 
quantities. However, in this naive approach infrared 
logarithmic divergences appear in the integrals.
By general arguments we then expect 
logarithmic corrections to the deduced scaling for $k_BT/\rho g \to \infty$. We can then only say that roughly  
\begin{eqnarray}
\frac{\hbar D V}{g} &\approx&  \frac{k_BT}{\rho g}  \label{eq:naive_D} \\
\frac{{\cal C}V}{\xi^3}  &\approx&  \left( \frac{k_BT}{\rho g} \right)^{1/2} \label{eq:naive_C}
\\
\frac{g \tau_c}{\hbar \xi^3}&\approx&  \left( \frac{k_BT}{\rho g} \right)^{-1/2} \label{eq:naive_tc}  \\
\frac{C_{\rm mc}(0) \hbar^2V \xi^3}{g^2}&\approx&  \left( \frac{k_BT}{\rho g} \right)^{3/2} \label{eq:naive_C(0)} \,,
\end{eqnarray}
where in (\ref{eq:naive_D})-(\ref{eq:naive_C(0)}) a factor $(k_BT/\rho g)^{3/2}$ comes from the Jacobian. 
A consequence of (\ref{eq:naive_D}) would be that at high temperature the diffusion coefficient is independent on $g$. 

This 
is compatible with the naive expectation that at high temperature the damping rate is proportional to the scattering cross section $\sigma=8\pi a^2 \propto g^2$, with a proportionality factor
independent of $a$ (as is the case for a classical gas where $\Gamma \simeq \rho \sigma v$
where $\rho$ is the density and $v$ the average velocity). In this naive
expectation, $\Gamma^{-1} \propto g^{-2}$ compensates
the contribution of $\vec{A}^2\propto g^2$, and $D$ is independent of $g$.
This is actually too naive and neglects logarithmic corrections.
For example, for the Landau damping rate, we were able to show
that, in the limit of a vanishing $\rho g$ for fixed temperature
$T$ and momentum $q$:
\begin{multline}
\check{\Gamma}_{\bf q}^L =
\frac{\pi}{q\xi} \frac{k_B T}{\rho g}
\left[
\ln\left(\frac{k_B T}{2\rho g}\right) + 
\ln\left(1-e^{-\beta \hbar^2 q^2/2m}\right) \right.\\
\left. + O\left(\frac{\rho g}{k_B T}\right)
\right].
\end{multline}

\section{Solution of the lossy classical field ergodic model}
\label{app:modele}

We first derive (\ref{eq:casT=0}), by solving the lossy ergodic
model exactly at zero temperature, 
and then derive (\ref{eq:raF}) for $T\neq0$, by solving the lossy ergodic model
to first order in the loss rate constant $\Gamma$.
After an explicit calculation we then obtain (\ref{eq:rap}).

To this end, we start with the fully quantum model,
defined by a Lindblad form master equation including one body losses,
and we use the formulation given in \cite{squeezing} 
of the Monte Carlo Wavefunction method for the expectation
value of an observable $\hat{O}$:
\begin{multline}
\langle \hat{O}\rangle(t) = \sum_{k\in \mathbb{N}}
\int_{0<t_1<\ldots<t_k<t} dt_1\ldots dt_k  \\
\sum_{\alpha_1,\ldots,\alpha_k}
\langle\psi(t)| \hat{O} |\psi(t)\rangle
\label{eq:reform}
\end{multline}
where the first sum is taken over the number $k$ of jumps, the
integrals are taken over the jump times $t_i$, the remaining sums
are taken over all possible types of jumps, and the $|\psi(t)\rangle$
is the unnormalized Monte Carlo wavefunction obtained from the initial
wavefunction by the deterministic non hermitian Hamiltonian evolution
interrupted at times $t_i$ by the action of the jump operators 
of type $\alpha_i$.
Here, for one body losses, the jump operators may be taken as
$C_{\mathbf{r}}=
dV^{1/2} \Gamma^{1/2} \hat{\psi}(\mathbf{r})$, where $\mathbf{r}$
is any point on the grid of the lattice model (of unit cell volume
$dV$). The jump associated to $C_{\mathbf{r}}$ then describes
the loss of a particle in point $\mathbf{r}$.
The non hermitian Hamiltonian is $H_{\rm eff}= H -
\frac{i\hbar}{2} \sum_{\mathbf{r}} C_{\mathbf{r}}^\dagger C_{\mathbf{r}}=
H - i\hbar\Gamma \hat{N}/2$, where $\hat{N}$ is the total number operator.

The lossy ergodic model is based on a classical field model,
where the state vector of the system is approximated by a Fock
state with $N(t)$ particles in the mode 
$\phi(\mathbf{r})$ linked to the classical field $\psi(\mathbf{r})$
by $\psi(\mathbf{r})=N^{1/2}(t) \phi(\mathbf{r})$.
Then the action of $C_{\mathbf{r}}$ on this Fock state
simply pulls out a factor $dV^{1/2} \Gamma^{1/2} \psi(\mathbf{r})$
in front of a Fock state with $N(t)-1$ particles in the mode $\phi$.
One may thus easily take the sum over the types of jumps in (\ref{eq:reform}),
$\alpha_i$ corresponding to a loss event in $\mathbf{r}_i$,
which produces factors equal to $\Gamma$ times
the updated atom number after successive jumps.
Also, in the lossy ergodic model, the condensate
accumulated phase $\varphi(t)$ is a classical quantity, evolving
with the rate $\dot\varphi(t) = -\mu_{\rm mc}[E(t),N(t)]/\hbar$.

At zero temperature, one then simply has $\dot\varphi(t)=-g N(t)/\hbar V$
so that, after a sequence of $k$ jumps at times $t_i$:
\begin{eqnarray}
\varphi(t) &=& -\frac{g}{\hbar V} [N t_1 + (N-1) (t_2-t_1) + \ldots
\nonumber \\
&&+ (N-k) (t-t_k)] \\
&=& -\frac{g}{\hbar V} \left[(N-k) t + \sum_{i=1}^{k} t_i\right]
\end{eqnarray}
where $N$ is here the initial atom number $N(0)$.
Similarly the squared norm of the Monte Carlo wavefunction 
after that sequence of jumps is
\begin{eqnarray}
\langle\psi(t)|\psi(t)\rangle &=& \Gamma^k N(N-1)\ldots (N-k+1)
\nonumber \\
&\times& e^{-\Gamma[N t_1 + \ldots + (N-k) (t-t_k)]} \\
&=& \Gamma^k \frac{N!}{(N-k)!} e^{-\Gamma[(N-k) t + \sum_{i=1}^{k} t_i]}.
\end{eqnarray}
The expectation value of $\varphi^n(t)$, $n$ positive integer, is thus
\begin{equation}
\langle \varphi^n(t) \rangle = 
\sum_{k=0}^{N} C_N^k \Gamma^k e^{-\Gamma (N-k) t} 
\int_{[0,t]^k}dt_1\ldots dt_k 
\varphi^n(t) \prod_{i=1}^{k} e^{-\Gamma t_i}
\end{equation}
where $C_N^k = N!/[k! (N-k)!]$ is the binomial
coefficient
and where we used the fact that the integrand was a symmetric function 
of the times $t_i$ to extend the integration domain to $[0,t]^k$
after division by $k!$.
After lengthy calculations, and using the values of the binomial
sums
\begin{equation}
\sum_{k=0}^{N} C_N^k e^{\lambda k} k^n = \partial_\lambda^n
(e^\lambda + 1)^N,
\end{equation}
we obtain the zero temperature results of the model:
\begin{eqnarray}
\langle \varphi(t)\rangle &=& -\frac{gN}{\hbar\Gamma V} 
\left(1-e^{-\Gamma t}\right) \\
\mbox{Var}\,\varphi(t) &=&  \left(\frac{g}{\hbar\Gamma V}\right)^2 N
\left(1-2\Gamma t\, e^{-\Gamma t} - e^{-2\Gamma t}\right).
\end{eqnarray}
This gives (\ref{eq:casT=0}).

Next, we solve the lossy ergodic model to first order in $\Gamma$,
at a non-zero temperature.
To this order, one can restrict to the contributions of the zero-jump
and of the single-jump trajectories. Calling $\mu$ the initial
microcanonical chemical potential, a function of the initial
(random) energy $E$ and (fixed) atom number $N$, and calling
$\mu+\delta \mu$ the value of the chemical potential after the first jump,
we have $-\hbar\varphi(t) = \mu t$ for the zero-jump trajectory
and $-\hbar\varphi(t) = \mu t_1 + (\mu+\delta\mu) (t-t_1) 
=\mu t + \delta\mu (t-t_1)$ for the single-jump trajectory
with a jump at time $t_1$. Thus 
\begin{multline}
\langle [-\hbar\varphi(t)]^n\rangle =
\langle (\mu t)^n \rangle e^{-\Gamma N t}
+ \int_0^t dt_1 \left\{e^{-\Gamma N t_1}\right. \\ 
\left. \times \Gamma N e^{-\Gamma(N-1)(t-t_1)}
\langle [\mu t + \delta\mu (t-t_1)]^n\rangle\right\} + O(\Gamma^2),
\end{multline}
where $\langle\ldots\rangle$ in the right hand side
stands for the expectation value over the initial
system energy $E$. To first order in $\Gamma$, the exponential
factors in the integral may be replaced by unity. Performing the
integral over $t_1$ gives
\begin{eqnarray}
\langle -\hbar\varphi(t)\rangle &=& \langle \mu\rangle t + 
\langle \delta\mu\rangle \frac{1}{2} \Gamma N t^2 + O(\Gamma^2) \\
\langle [-\hbar\varphi(t)]^2\rangle &=& \langle \mu^2\rangle t^2 + 
\langle (\delta\mu)^2\rangle \frac{1}{3} \Gamma N t^3\nonumber \\
&+& \langle \mu \delta\mu\rangle \Gamma N t^3 
+O(\Gamma^2).
\end{eqnarray}
This leads to (\ref{eq:raF}).
Note that the final result here is valid for $\Gamma t \ll 1$, even
if our derivation seems to request the stronger condition $\Gamma N t \ll 1$.

Explicit expressions may be obtained from 
(\ref{eq:E'}),(\ref{eq:N'}),(\ref{eq:itcfm1}),(\ref{eq:itcfm2}),(\ref{eq:itcfm3}), 
and in the thermodynamic
limit, where in particular one may approximate
$f(N-1)$ by $f(N) - df(N)/dN$.
Setting
\begin{eqnarray}
S(N) &=& \frac{1}{\mathcal{M}} \sum_{\mathbf{k}\neq\mathbf{0}}
\left(\frac{\hbar^2 k^2}{2m} + \frac{2gN}{V}\right)^{-1} \\
\Sigma(N) &=& \frac{dS}{dN}(N) + \frac{g}{V} S^2(N) +
\frac{1}{N} S(N),
\end{eqnarray}
and noting that $S(N) = O(V^0)$ and $\Sigma(N)= O(1/V)$ in the thermodynamic
limit, we have
\begin{eqnarray}
\mu &=& \frac{g}{V}[N+(E-E_0) S(N)] \\
\delta E- \delta E_0 &=& -\frac{E-E_0}{N} - (\mu-\mu_0) + O(V^{-1}) \\
\delta\mu &=&  -\frac{g}{V} [1+(E-E_0) \Sigma(N)]  \nonumber \\
&+& O(V^{-2})
\end{eqnarray}
where $\delta E$ is the energy change after the first jump
and $\mu_0 = g N/V$ is the zero temperature (classical field) chemical
potential.
Taking the expectation value in (\ref{eq:raF}) over the initial
system energy $E$ gives
\begin{multline}
\mbox{Var}\, \hbar\varphi(t) \simeq
\left(\frac{g}{V}\right)^2 S^2(N) t^2 \mbox{Var}\, (E-E_0) \\
+ \frac{1}{3} \Gamma N t^3 \left(\frac{g}{V}\right)^2 
\left[1
-3 S(N) \Sigma(N) \mbox{Var}\, (E -E_0)
\right.  \\
+ \left.
2\langle E-E_0\rangle \Sigma(N)
+\langle E-E_0\rangle^2 \Sigma^2(N)
\right].
\end{multline}
Here one simply has $\mbox{Var}\, (E-E_0)=\mbox{Var}\,E$ since
the initial particle number is fixed so that the ground state energy
$E_0$ does not fluctuate. For a classical field model
in the canonical ensemble, $\langle E-E_0\rangle = \mathcal{M} k_B T$
and $\mbox{Var}\, (E-E_0) = \mathcal{M} (k_B T)^2$.

In the limit $k_B T \gg N g/V$, which is natural for a classical field
model, the above expression for $\mbox{Var}\, \hbar\varphi(t) $
may be greatly simplified. Taking a momentum cut-off $K$
such that $\hbar^2 K^2/2m = k_B T$, and ignoring numerical factors,
we obtain in the thermodynamical limit and high temperature limit:
\begin{eqnarray}
\mathcal{M} & \approx & V K^3 \\
S(N) & \approx & \frac{1}{k_B T} \\
N \frac{dS}{dN}(N) &\approx& - \frac{S(N)}{K\xi} \ll S(N) \\
\frac{Ng}{V} S^2(N) &\approx& \frac{S(N)}{(K\xi)^2} \ll S(N)  \\
\Sigma(N) &\approx& \frac{S(N)}{N} \approx \frac{1}{Nk_B T} \\
S(N) \Sigma(N) \mbox{Var}\, (E -E_0) &\approx& \frac{K^3}{\rho} \\
\langle E-E_0\rangle \Sigma(N) &\approx& \frac{K^3}{\rho}.
\end{eqnarray}
Since $K^3/\rho$ is of the order of the non condensed fraction
$\langle \delta N\rangle/N$,  supposed to be $\ll 1$ here,
we recover (\ref{eq:rap}).

\section{Quantum single mode model with one body losses}
\label{app:modeleQ}

We show here that (\ref{eq:casT=0}), obtained at zero temperature within
a classical field model, extends to the quantum case of a pure condensate
with a large atom number and in an initial number state with $N$
particles.

The master equation for the single mode quantum model density operator
$\hat{\rho}$ with one body losses is
\begin{equation}
\frac{d\hat{\rho}}{dt} = \frac{1}{i\hbar} [\hat{H},\hat{\rho}] 
+\Gamma \hat{a}_0 \hat{\rho} \hat{a}_0^\dagger
-\frac{\Gamma}{2} \{ \hat{a}_0^\dagger \hat{a}_0,\hat{\rho}\}
\label{eq:ME}
\end{equation}
where $\hat{a}_0$ annihilates a particle in the condensate mode,
and $\hat{H}=g \hat{a}_0^{\dagger 2}\hat{a}_0^2 /(2V)$.
A useful consequence is that the mean value of a not explicitly time
dependent operator $\hat{O}$ evolves as
\begin{multline}
\frac{d}{dt} \langle \hat{O}\rangle \equiv 
\frac{d}{dt} \mbox{Tr} [\hat{O} \hat{\rho}] 
= \langle \frac{1}{i\hbar} [\hat{O}, \hat{H}]\rangle \\
+\frac{\Gamma}{2} \langle
\hat{a}_0^\dagger [\hat{O},\hat{a}_0] +
[\hat{a}_0^\dagger, \hat{O}] \hat{a}_0\rangle.
\label{eq:utile}
\end{multline}

Neglecting the possibility that the condensate mode is empty,
we use the modulus-phase representation $\hat{a}_0 = e^{i\hat{\theta}}
\hat{N}_0^{1/2}$ where the phase operator $\hat{\theta}$ and the number
operator $\hat{N}_0=\hat{a}_0^\dagger\hat{a}_0$ obey
the commutation relation $[\hat{\theta},\hat{N}_0]=-i$.
In Heisenberg picture, the incremental evolution of the phase operator
during an infinitesimal time step $dt$ involves, in addition to
the usual commutator with the Hamiltonian $\hat{H}$, a deterministic
term $\hat{A}$ and a quantum stochastic term $d\hat{B}$ 
scaling as $dt^{1/2}$
due to the losses \cite{Book}:
\begin{equation}
d\hat{\theta} = \frac{dt}{i\hbar} [\hat{\theta},H] + \hat{A}\, dt
+ d\hat{B}.
\end{equation}
Applying (\ref{eq:utile}) to $\hat{O}=\hat{\theta}$ gives
\begin{equation}
\hat{A} \equiv 0.
\end{equation}
Applying (\ref{eq:utile}) to $\hat{O}=\hat{\theta}^2$ gives
\begin{equation}
\langle \overline{d\hat{B}^2}\rangle  = \Gamma dt \langle
[\hat{a}_0^\dagger,\hat{\theta}]
[\hat{\theta},\hat{a}_0]\rangle = \langle \frac{\Gamma dt}{4\hat{N}_0}\rangle.
\end{equation}
In the large occupation number limit, we may thus neglect $d\hat{B}$
and take 
\begin{equation}
\frac{d}{dt}\hat{\theta} \simeq [\hat{\theta},\hat{H}]/i\hbar = 
-g(\hat{N}_0-1/2)/\hbar V.
\end{equation}
This justifies the assumption in the classical field model
that the condensate phase is not affected by a jump.
In the quantum model, the variance of the
condensate accumulated phase $\hat{\varphi}(t)=\int_0^t d\tau
[d\hat{\theta}(\tau)/d\tau]$ is thus
\begin{multline}
\mbox{Var}\, \hat{\varphi}(t) \simeq \left(\frac{g}{\hbar V}\right)^2 
\int_0^t d\tau \int_0^t d\tau'
\left[\langle\hat{N}_0(\tau)\hat{N}_0(\tau')\rangle \right.
\\ -
\left. \langle\hat{N}_0(\tau)\rangle \langle\hat{N}_0(\tau')\rangle\right].
\label{eq:pfin}
\end{multline}

To calculate the one time averages of $\hat{N}_0$ and $\hat{N}_0^2$
we use (\ref{eq:utile}) with $\hat{O}=\hat{N}_0$
and $\hat{O}=\hat{N}_0^2$:
\begin{eqnarray}
d\langle\hat{N}_0\rangle/dt & =& - \Gamma \langle\hat{N}_0\rangle \\
d\langle\hat{N}_0^2\rangle/dt & =& - 2\Gamma \langle\hat{N}_0^2\rangle
+ \Gamma \langle\hat{N}_0\rangle
\end{eqnarray}
that are straightforward to integrate with initial conditions
$\langle\hat{N}_0(t=0)\rangle=N$ and
$\langle\hat{N}_0^2(t=0)\rangle=N^2$.

To calculate the two time averages, we can restrict to $\tau\geq \tau'$
by hermitian conjugation. Then we use the quantum regression theorem:
setting $\hat{\sigma}(\tau') = \hat{N}_0(0) \hat{\rho}(\tau')$,
the ``density operator" $\hat{\sigma}(\tau)$ evolves at later times $\tau
\geq \tau'$
with the same master equation as $\hat{\rho}$, and
\begin{equation}
\langle \hat{N}_0(\tau) \hat{N}_0 (\tau')\rangle = \mbox{Tr}
[\hat{N}_0(0) \hat{\sigma}(\tau)]
\end{equation}
for $\tau \geq \tau'$. As a consequence
\begin{equation}
\frac{d}{d\tau} \langle \hat{N}_0(\tau) \hat{N}_0 (\tau')\rangle =
-\Gamma \langle \hat{N}_0(\tau) \hat{N}_0 (\tau')\rangle
\end{equation}
for $\tau \geq \tau'$, which is straightforward to integrate
with the initial condition at $\tau=\tau'$,  
$\langle \hat{N}_0^2(\tau')\rangle$.
We obtain for $\tau\geq \tau'\geq 0$, and for an initial number state
with $N$ particles:
\begin{eqnarray}
\langle \hat{N}_0(\tau)\rangle &=& N e^{-\Gamma \tau} \\
\langle \hat{N}_0^2(\tau)\rangle &=& N^2 e^{-2\Gamma \tau} +
N e^{-\Gamma \tau} \left(1-e^{-\Gamma \tau}\right)\\
\langle \hat{N}_0(\tau) \hat{N}_0 (\tau')\rangle &=& e^{-\Gamma(\tau-\tau')}
\langle \hat{N}_0^2(\tau')\rangle.
\end{eqnarray}
Mapping the double integral in (\ref{eq:pfin}) to the
integration domain $0\leq \tau'\leq \tau$ leads to 
\begin{equation}
{\mbox{Var}} \, \hat{\varphi}(t) \simeq 
\left( \frac{g}{V\hbar \Gamma}\right)^2 N \left[ 1 - 2\Gamma t \, e^{-\Gamma t} - e^{-2\Gamma t}\right] \,,
\end{equation}
which coincides with the zero temperature classical field
model result (\ref{eq:casT=0}).


\begin{thebibliography}{99}
\bibitem{note_phase}
At a temperature $T$ much below the critical temperature $T_c$
and for weak interactions, we can neglect the possibility 
that the condensate is empty, and we can introduce 
the modulus-phase representation of the annihilation operator of
a condensate particle $\hat{a}_0$, 
$\hat{a}_0=e^{i\hat{\theta}} \sqrt{\hat{N}_0}$,
where the operator $\hat{\theta}$ represents the condensate phase
and $\hat{N}_0$,
the number of condensate particles, is its conjugate variable so that
$[\hat{N}_0,\hat{\theta}]=i$.
\bibitem{note_corr_symm} In general one defines a symmetrized correlation function $C_s(t)=(C(t)+C(t)^\ast)/2$. By construction, in stationary conditions $C_s(t)=C_s(-t)$. In our approach however, the correlation function $C(t)$ will be real so that $C_s(t)=C(t)$.
\bibitem{note_cutoff} The highest Bogoliubov energy on the cubic lattice is 
$\epsilon_{k_{\rm max}}=[(\hbar^2 k_{\rm max}^2/2m)(\hbar^2 k_{\rm max}^2/2m+2\rho g)]^{1/2}=k_BT$.
\bibitem{t_tc_only} For $k_BT \gg \rho g$, the quantity $\tilde{{\cal A}}(\rho a^3)^{1/2}$ does not depend on the interaction strength $g$ and 
is proportional to the fraction of non condensed particles, scaling here as $(T/T_c)^{3/2}$: 
$\tilde{{\cal A}}(\rho a^3)^{1/2}=\frac{(\pi/2)^{1/2}12 \zeta(3/2)}{5 \zeta(5/2)}(T/T_c)^{3/2}
\simeq 5.86 (T/T_c)^{3/2}$.
\bibitem{one_can_show}
From (\ref{eq:def_alpha}) and (\ref{eq:Qcan}) it is apparent that
$Q^{\rm can}(0) \vec{\epsilon}$ is proportional to the vector
$\vec{\alpha}$. As a consequence, $P^\dagger Q^{\rm can}(0)
\vec{\epsilon}$ is equal to zero, so that
$P^\dagger Q^{\rm can}(0) P = P^\dagger Q^{\rm can}(0)$. 
Since $\vec{X}_{\rm can}(0)= P^\dagger Q^{\rm can}(0) \vec{A}$,
and $\vec{X}_{\rm mc}(0)= P^\dagger Q^{\rm can}(0) P \vec{A}$,
this implies that $\vec{X}_{\rm can}(0)= \vec{X}_{\rm mc}(0)$.
\bibitem{note_cin_int}
In order to calculate the energy change one has in principle 
to know $E_{\rm  kin}$ and $E_{\rm  int}$ separately. Since $E_{\rm  kin}$ and 
$E_{\rm  int}$ are not constants of motion, this means that one has to 
know the field $\psi$. In (\ref{eq:E'}), taking the microcanonical averages, 
we in fact neglect the fluctuations of $E'-E$. This is consistent with the 
fact that in (\ref{eq:thdot_mod}) we neglect the fluctuations of $\dot{\theta}$
and replace $\dot{\theta}$ with its microcanonical average. As a consequence
with this model we do not address the diffusive term in the phase spreading.
\end{thebibliography}
\end{document}